\documentclass[a4paper,reqno, 11pt]{amsart}

\usepackage{amsrefs}
\usepackage{amsfonts}
\usepackage{amssymb}
\usepackage{amscd}
\usepackage{fullpage}
\usepackage[labelformat=empty]{subfig}
\usepackage{booktabs}
\usepackage{xypic}
\usepackage[dvips]{graphicx,epsfig}
\usepackage{enumerate}

\BibSpec{article}{%
+{}{\sc \PrintAuthors} {author}
+{,}{ \textrm} {title}
+{,}{ \textit} {journal}
+{,}{ } {volume}
+{}{ \IfEmptyBibField{journal}{}{\PrintDate{date}}} {transition}
+{,}{ } {pages}
+{,}{ } {status}
+{.}{} {transition}
+{}{ Preprint\IfEmptyBibField{date}{}{ \PrintDate{date}}, available at \tt} {eprint}
+{.}{} {transition}
}

\BibSpec{book}{%
+{}{\sc \PrintAuthors} {author}
+{,}{ \textit} {title}
+{.}{ } {series}
+{,}{ } {volume}
+{.}{ } {publisher}
+{,}{ } {place}
+{,}{ } {date}
+{.}{} {transition}
}

\BibSpec{collection.article}{
+{} {\sc \PrintAuthors} {author}
+{,} { \textrm } {title}
+{,} { in \it \PrintConference} {conference}
+{,} { pp.~} {pages}
+{.} {\PrintBook} {book}
+{,} { \PrintDateB} {date}
+{.} {} {transition}
}

\BibSpec{innerbook}{
+{.} { \emph} {title}
+{.} { } {part}
+{:} { \emph} {subtitle}
+{.} { } {series}
+{,} { } {volume}
+{.} { Edited by \PrintNameList} {editor}
+{.} { Translated by \PrintNameList}{translator}
+{.} { \PrintContributions} {contribution}
+{.} { } {publisher}
+{.} { } {organization}
+{,} { } {address}
+{,} { \PrintEdition} {edition}
+{,} { \PrintDateB} {date}
+{.} { } {note}
}

\allowdisplaybreaks[3]

\newcommand{\de}{{\partial}}

\newcommand{\bbN}{\mathbb{N}}
\newcommand{\bbK}{\mathbb{K}}

\newcommand{\bbZ}{\mathbb{Z}}
\newcommand{\bbR}{\mathbb{R}}
\newcommand{\bbC}{\mathbb{C}}
\newcommand{\bbP}{\mathbb{P}}

\def\bary{\begin{array}} 
\def\eary{\end{array}} 
\def\ben{\begin{enumerate}} 
\def\een{\end{enumerate}}
\def\bit{\begin{itemize}} 
\def\eit{\end{itemize}}
\def\nn{\nonumber} 

\newcommand{\cO}{\mathcal{O}}

\newcommand{\cC}{\mathcal{C}}

\newcommand{\LL}{\mathcal{L}}

\newcommand{\cN}{\mathcal{N}}

\newcommand{\cF}{\mathcal{F}}

\newcommand{\cV}{\mathcal{V}}

\newcommand{\cM}{\mathcal M}

\newcommand{\bfz}{\mathbf{z}}
\newcommand{\bfu}{\mathbf{u}}
\def\beq{\begin{equation}}                     %
\def\eeq{\end{equation}}                       %
\def\bea{\begin{eqnarray}}                     
\def\eea{\end{eqnarray}}
\def\bary{\begin{array}} 
\def\eary{\end{array}} 
\def\ben{\begin{enumerate}} 
\def\een{\end{enumerate}}
\def\bit{\begin{itemize}} 
\def\eit{\end{itemize}}
\def\nn{\nonumber} 
\def\de {\partial}

\def\a{\alpha}
\def\b{\beta}
\def\g{\gamma}
\def\d{\delta}
\def\rd{\mathrm{d}}
\def\e{\epsilon}

\theoremstyle{plain}

\newtheorem{thm}{Theorem}[section]
\newtheorem{lem}[thm]{Lemma}
\newtheorem{assm}{Assumption}[section]
\newtheorem{prop}[thm]{Proposition}
\newtheorem{conj}[thm]{Conjecture}

\newtheorem*{conj*}{Conjecture}

\newtheorem*{cor*}{Corollary}

\newtheorem{rem}[thm]{Remark}
\newtheorem{defn}{Definition}[section]

\theoremstyle{definition}

\newcommand{\Li}{\operatorname{Li}}
\newcommand{\Snp}{\operatorname{S}}

\newcommand{\GIT}[1]{/\!\!/_{\kern-.2em #1 \kern0.1em}}

\renewcommand{\l}{\left}
\renewcommand{\r}{\right}
\newcommand{\bra}{\left\langle}
\newcommand{\ket}{\right\rangle}
\newcommand{\pf}{\noindent {\it Proof.} }

\newcommand{\ev}{\operatorname{ev}}

\begin{document}

\title[The local Gromov--Witten theory of $\bbC \bbP^1$ and integrable hierarchies]{The local Gromov--Witten theory of $\bbC \bbP^1$ and integrable hierarchies}

\author{Andrea Brini}
\address{Section de Math\'ematiques, Universit\'e de Gen\`eve, 2-4 Rue du Li\`evre 1210, Geneva, Switzerland}
\email{Andrea.Brini@unige.ch}

\begin{abstract}

In this paper we begin the study of the relationship between the local
Gromov--Witten theory of Calabi--Yau rank two bundles over the projective line
and the theory of integrable hierarchies. We first of all construct
explicitly, in a large number of cases, the Hamiltonian dispersionless
hierarchies that govern the full-descendent genus zero theory. Our main tool
is the application of Dubrovin's formalism, based on associativity equations,
to the known results on the genus zero theory from local mirror symmetry and
localization. The hierarchies we find are apparently new, with the exception
of the resolved conifold $\cO_{\bbP^1}(-1) \oplus \cO_{\bbP^1}(-1)$ in the
equivariantly Calabi--Yau case. For this example the relevant dispersionless
system turns out to be related to the long-wave limit of the  Ablowitz--Ladik lattice. This identification provides us with a complete procedure to reconstruct the dispersive hierarchy which should conjecturally be related to the higher genus theory of the resolved conifold. We give a complete proof of this conjecture for genus $g\leq 1$; our methods are based on establishing, analogously to the case of KdV, a ``quasi-triviality'' property for the Ablowitz--Ladik hierarchy at the leading order of the dispersive expansion. We furthermore  provide compelling evidence in favour of the resolved conifold/Ablowitz--Ladik correspondence at higher genus by testing it successfully in the primary sector for $g=2$.

\end{abstract}

\maketitle
\tableofcontents

\clearpage
\section{Introduction}
\subsection{Gromov--Witten invariants and integrable hierarchies}
\label{sec:introgen}
Gromov--Witten theory deals with the study and the computation of intersection numbers on moduli spaces of holomorphic maps from a source Riemann surface to a compact K\"ahler manifold $X$. Denote by $\overline{\cM}_{g,n}(X,\b)$ the Kontsevich compactification of the moduli space of degree $\b \in H_2(X,\bbZ)$ stable maps from $n$-pointed genus $g$ curves to $X$. The Gromov--Witten invariants of $X$ are defined as
\beq
\bra \tau_{p_1}(\phi_{\a_1}) \dots  \tau_{p_n}(\phi_{\a_n}) \ket_{g,n,\b}^X :=
\int_{[\overline{\cM}_{g,n}(X,\b)]^{\rm vir}} \prod_{i=1}^n \ev^*_i(\phi_{\a_i})\psi_i^{p_i},
\eeq
where $[\overline{\cM}_{g,n}(X,\b)]^{\rm vir}$ is the virtual fundamental
class of $\overline{\cM}_{g,n}(X,\b)$, $\phi_{\a_i}\in H^\bullet(X, \bbC)$ are
arbitrary co-homology classes of $X$,  $\ev_i:\overline{\cM}_{g,n}(X,\b)\to X$
is the evaluation map at the $i^{\rm th}$ marked point, and $\psi_i=c_1(\LL_i)$ are the first Chern classes of the universal cotangent line bundles $\LL_i$ on $\overline{\cM}_{g,n}(X,\b)$. When $p_i=0$ for all $i$, these invariants have an interpretation as a ``count'' (in a suitable sense) of holomorphic curves of genus $g$ and degree $\b$ inside $X$, subject to the constraint of intersecting $n$ generic cycles given by the Poincar\'e duals of the classes $\phi_{\a_i}$. \\

We know from examples
 \cites{Witten:1990hr, Kontsevich:1992ti, MR2199226},
and have limited general evidence both in the Fano and the Calabi--Yau case
\cites{Eguchi:1997jd, MR1901075, MR2208418, Bershadsky:1993cx}, 
that Gromov--Witten invariants of a target space $X$ could be subject to a mysterious web of constraints relating them to one another,  and a long-standing problem in the subject has been to lift at least part of the mystery. An influential conjecture 
 stemming from Witten's influential work on two-dimensional topological
 gravity \cite{Witten:1990hr} asserts that a full explanation should be
 provided by the theory of integrable hierarchies of non-linear PDEs. More
 precisely, introduce formal symbols $\e$ and $t^{\alpha,p}$, where $\a \in
 \{1,\dots, h_X\}$, $h_X :=\dim_\bbC H^\bullet(X, \bbC)$ and $p \in \bbN$; the
 set $\l\{t^{\a,p}\r\}_{\substack{\a \in \{1, \dots, h_X\} \\ p\in \bbN}}$ will be shorthandedly written as $\mathbf{t}$. Moreover let $\phi_1$ correspond to the unity of $H^\bullet(X)$ and denote the formal variable $t^{1,0}$ with $x$. We define the all-genus, full-descendent Gromov--Witten potential of $X$ as the formal power series

\bea
 \mathcal{F}^X(\e, \mathbf{t}) = \sum_{g \geq 0} \e^{2g-2}\sum_{ \beta \in
   H_2(X,\bbZ)}\sum_{n \geq 0} \sum_{\substack{\a_1, \ldots, \a_n \\p_1,\ldots,p_n}}
   {\prod_{i=1}^n t^{\a_i,p_i} \over n!}
\bra \tau_{p_1}(\phi_{\a_1}) \dots  \tau_{p_n}(\phi_{\a_n}) \ket_{g,n,\b}^X.
\label{eq:pot}
\eea

\noindent The Gromov--Witten/Integrable Systems correspondence can then be stated as follows:

\begin{conj}
\label{conj:integrableGW}
Let $\cF^X(\e, \mathbf{t})$ denote the all-genus full descendent Gromov--Witten potential of $X$. Then there exists a Hamiltonian integrable hierarchy of PDEs such that  $\mathcal{F}^X(\e, \mathbf{t})$ is the logarithm of a $\tau$--function associated to one of its solutions. The variables $t^{\a,p}$ are identified with times of the hierarchy, and the genus counting variable $\e$ with a perturbative parameter in a small dispersion expansion of the equations.
\end{conj}
\noindent By ``small dispersion expansion'' we mean that, in terms of the basic fields $u_\a(\mathbf{t})$ 
\beq
u_\a(\mathbf{t}):=\epsilon^2 \frac{\de^2 \cF^X(\e, \mathbf{t})}{\de x \de t^{\a,0}},
\label{eq:tau}
\eeq
the equations of the hierarchy should take the form of a formal gradient expansion
\beq
\frac{\de u_\a}{\de t^{b,p}}= \sum_{g=0} \e^{2g}
\sum_{\beta=1}^{h_X}A^{[g]}(\mathbf{u}, \mathbf{u}_x,  \mathbf{u}_{xx},  \ldots,\mathbf{u}^{(2g+1)} ).
\label{eq:dispexp}
\eeq
In \eqref{eq:dispexp} $A^{[g]}$ are degree $2g+1$ homogeneous polynomials in $\mathbf{u}^{(n)}$, where we have defined
\beq
 \deg \frac{\de^n u_\alpha}{\de x^n}=n \quad \forall \a.
\label{eq:grading}
\eeq

While the existence of {\it some} tau-symmetric integrable system associated
to the Gromov--Witten theory of a given target space $X$
follows basically\footnote{Except for the subtle issue of polynomiality of the flows;
  see \cite{2010arXiv1009.5351B} for recent progress in this direction.}
from the $3g-2$ theorem \cite{Eguchi:1998ji, MR2115766}, a {\it constructive} proof of this conjecture - i.e., an explicit characterization of the hierarchy - would be a 
far-reaching result, both in principle and computationally.
%
However, to find out whether such an integrable structure can be found and effectively described is in general a tough task, and 
the catalogue of rigorous and complete answers to this question is restricted to a discouragingly low number of examples:

\ben
\item $X=\mathrm{pt}$, that is, intersection theory on the Deligne--Mumford compactification of the moduli space of curves. The Witten--Kontsevich theorem states \cite{Witten:1990hr,Kontsevich:1992ti} that the KdV hierarchy is the relevant integrable system in this case;
\item $X=\bbP^1$, in which case the associated system is the extended Toda hierarchy \cite{Eguchi:1994in,  MR2092034, MR2199226, Milanov:2006cu, MR2309158};
\item $X=(\bbP^1)^{\circlearrowleft T\simeq \bbC^*}$, where $T$ is the canonical torus action on $\bbP^1$.  The relevant hierarchy is a reduction of 2D-Toda \cite{MR2049645, MR2439568, MR2199226}.
\een
For each of the three cases above, a few proposals have been made to extend the correspondence to orbifolds of the form $[X/G]$, where $G$ is a finite group \cite{MR1950944, rossi-2008, MR2433616, pauljohnsonthesis}; the corresponding candidate hierarchies should be reductions of KP (resp. 2D-Toda) for $X=\mathrm{pt}$ (resp. $X=\bbP^1$). Unfortunately, apart from this very limited bestiary, the goal to have a general constructive proof of Conjecture \ref{conj:integrableGW} appears to be out of reach at the moment. In fact, even adding new examples to the above list seems to be a very challenging  problem: the next-to-simplest case of the complex projective plane $\bbP^2$ is already hard to tackle, and it is as of today unsolved. \\

On the other hand, recent developments \cite{Aganagic:2003db, Aganagic:2003qj}
strongly indicate a natural new arena to push forward the study of the
Gromov--Witten/Integrable Systems correspondence: the local theory of toric
Calabi--Yau threefolds. In this context, physics-inspired dualities
 have provided an
impressive quantity of new insights, including conjectural proposals for the solutions of the
non-equivariant theory \cite{Aganagic:2003db,
  Bouchard:2007ys} and remarkable connections to other areas of Mathematics: examples include
other moduli space problems in Algebraic Geometry \cite{MR2264664, moop} and quite different subjects like quantum topology
\cite{Gopakumar:1998ki, Ooguri:1999bv, Marino:2001re} and modular forms
\cite{Aganagic:2006wq}. On one hand, it is natural to speculate that the high degree of
solvability of the theory could be {\it explained} by underlying integrable
structures; on the other, such a rich web of mathematical interconnections renders the possibility to elucidate
the role of integrability in this context an even more appealing goal. \\

\subsection{Main results}
\label{sec:introresults}
In this paper we begin to address this problem by studying the integrable structures that govern the equivariant Gromov--Witten theory of Calabi--Yau rank two bundles over the complex projective line - that is, differential neighbourhoods of a (not necessarily isolated) rational curve inside a Calabi--Yau threefold. By Grothendieck's theorem, such bundles split 
into a sum of line bundles: $\mathcal{O}_{\mathbb{P}^1}(n_1)\oplus \mathcal{O}_{\mathbb{P}^1}(n_2)$, $n_i \in \mathbb{Z}$; by the Calabi--Yau condition, we must have that $k:=-n_1=n_2+2$. We will denote by $X_k$ the total spaces of these bundles. Moreover, we will consider their equivariant Gromov--Witten theory with respect to a $T\simeq \mathbb{C}^*$ torus action, which covers the trivial action on the base $\bbP^1$ and rotates the fibers:
\beq 
X_k := \cO_{\bbP^1}(-k)\oplus \cO_{\bbP^1}(k-2)^{\circlearrowleft T}, \quad k\in \bbZ.
\label{eq:Xk}
\eeq
In many cases, we will take $T$ to act with identical (resp. opposite) characters on the two fibers; we will refer to these choices as the {\it diagonal} (resp. {\it anti-diagonal}) case. \\

In spirit, our study will be very close to the perturbative philosophy of Dubrovin--Zhang \cite{dubrovin-2001} for the non-equivariant Gromov--Witten theory of Fano manifolds with $(p,p)$ co-homology\footnote{This represents the prototypical  family of target spaces whose big quantum co-homologies satisfy  the technical assumptions necessary for Dubrovin and Zhang's machinery to work, like commutativity and semi-simplicity of the quantum product and a well-defined grading.}.
Let us briefly recall the main lines of their strategy. In their case, the whole hierarchy is constructed according to the following two-step process:
\ben
\item find a closed form description of its genus zero approximation (the {\it Principal Hierarchy});
\item find a reconstruction procedure to incorporate the higher genus corrections.
\een
Step (1) is based on the datum of a Frobenius manifold, that is, a solution of the Witten--Dijkgraaf--Verlinde--Verlinde equations possessing a distinguished dependence on one of its variables (the unity direction) and obeying a quasi-homogeneity condition (existence of the Euler field). Out of these data, it was shown in \cite{Dubrovin:1992dz} how  to associate a quasi-linear, non-dispersive Hamiltonian hierarchy and a $\tau$-function coinciding with the genus zero Gromov--Witten partition function. Step (2) is much more involved, and strongly relies on the  the existence of a local bi-Hamiltonian structure, as well as on the assumption of semi-simplicity of the quantum product and of Virasoro constraints on the dispersionful $\tau$-function \cite{Dubrovin:1997bv, dubrovin-2001}. \\

We will try to transfer some of the guiding principles of \cite{dubrovin-2001}
to the case at hand. A major obstacle is the fact that equivariant quantum
co-homology rings do not satisfy all axioms of a Frobenius manifold, and in particular the quasi-conformality of the prepotential. Still, the arguments of \cite{Dubrovin:1992dz} show that Step (1) above is {\it almost independent} of the presence of an Euler vector field, the only requirement being that the prepotential be known in closed form. In other words, bi-Hamiltonianity is not required to reconstruct the Principal Hierarchy; the existence of a grading operator is only needed to fix completely a canonical basis of flows. \\
For the case of the local theory of $\bbP^1$ in the diagonal and anti-diagonal case, and for the resolved conifold with a generic $(\bbC^*)^2$-fiberwise action, we have complete control on the prepotential both from the $A$-model \cite{Bryan:2004iq} and the $B$-model side \cite{MR2276766}. This will be sufficient for us to construct in a completely explicit way the relevant tree-level hierarchies.  \\

From a geometer's point of view, however, the real utility of a clear link with integrable hierarchies resides in the possibility to effectively perform Step (2), namely, to give a complete recipe to solve the all-genus, full-descendent theory in terms of a dispersive deformation of the Principal Hierarchy.  This would be particularly valuable for the case at hand, where little is known about possible higher genus relations between descendent invariants. For this second step, however, it looks hopeless to generalize the methods of Dubrovin--Zhang for the construction of the dispersive tail, as the validity of some of their key assumptions, like existence of Virasoro symmetries annihilating the $\tau$-function, is  unclear, if not in jeopardy in our case. \\

Still, in one example we can find a way out. It turns out that for the resolved conifold $\cO_{\bbP^1}(-1)\oplus \cO_{\bbP^1}(-1)$ with anti-diagonal action the Principal Hierarchy coincides with the long-wave limit of the so-called Ablowitz--Ladik lattice \cite{MR0377223}. The latter can be regarded as a complexified version of the discretized non-linear Schr\"odinger hierarchy, and appears as a particular reduction of the 2D-Toda hierarchy. Explicit knowledge of a candidate dispersive integrable model allows us to give a full reconstruction of the dispersive flows. It is tempting to speculate that this particular deformation could be the one that verifies Conjecture \ref{conj:integrableGW} in this case.

\begin{conj}
The all-genus, full descendent Gromov--Witten potential of the resolved conifold $X_{1}$ in the equivariantly Calabi--Yau case is the logarithm of a $\tau$-function of the Ablowitz--Ladik hierarchy.
\label{conj:AL}
\end{conj}

If proven, this statement would add a fourth item to the list we presented in
Sec. \ref{sec:introgen}. We have various reasons to believe that this
conjecture is true. First of all, it was shown in \cite{MR2462355} that, for
$2$-component integrable systems like the ones we consider in this paper,
integrability very often breaks down when we turn on dispersive
perturbations. This is for example the case of the generalized Fermi--Pasta--Ulam systems, for which the procedure of discretizing space derivatives never preserves involutivity of the flows except for exponential non-linearities (i.e. for the Toda lattice). Having {\it one} dispersive integrable candidate is an already fortunate circumstance and, if we trust the statement of Conjecture \ref{conj:integrableGW}, it should be taken very seriously. \\ The second, and much more cogent piece of evidence that we provide is given by the following
\begin{thm}
Conjecture \ref{conj:AL} is true for $g\leq 1$.
\label{thm:main}
\end{thm}
The key idea in our proof will be to establish a so-called
``quasi-triviality'' property for the Ablowitz--Ladik hierarchy at the leading
order of the dispersive expansion. Although differing in the way we obtained
it, due to the apparent absence of a second compatible local Poisson bracket for the Ablowitz--Ladik system, our final result comes very close to analogous statements in the bi-Hamiltonian case \cite{Dubrovin:1997bv, dubrovin-2001}.\\ Finally, we will exploit the possibility to reconstruct the dispersive flows, order by order in the parameter $\epsilon$, to give higher genus tests of our proposal. In particular we verify the following non-trivial implication of Conjecture \ref{conj:AL}:
\begin{thm}
Under Assumption \ref{assm:tausymm} (see Sec.~\ref{sec:g2}), let $\cF(\e,\mathbf{t})$ be the Ablowitz--Ladik $\tau$-function which reduces for $\epsilon \to 0$ to the topological $\tau$-function of the Principal Hierarchy of $X_{1}$ with anti-diagonal action. Then its reduction to small phase space at $\cO(\epsilon^4)$ coincides with the genus $2$ primary Gromov--Witten potential of the resolved conifold in the equivariantly Calabi--Yau case, possibly up to the degree zero term.
\label{cor:g=2}
\end{thm}
For the reasons that we have outlined at the end of section
\ref{sec:introgen}, we believe that this new example of the
Gromov--Witten/Integrable Systems correspondence could be a good starting point
for new insights in toric Gromov--Witten theory. A partial list of the
questions to be answered include the relationship of our hierarchies with the
physicists' open invariants of toric Calabi--Yau threefolds and the
Eynard--Orantin recursion \cite{Eynard:2007kz, Bouchard:2007ys}, a local mirror
symmetry description of the hierarchies in the framework of spectral curves
and the universal Whitham hierarchy \cite{Krichever:1992qe, Aganagic:2003qj},
the study of the  fate of the Virasoro conjecture \cite{Eguchi:1997jd} in the
equivariant case, 
multi-parameter generalizations ({\it e.g.} the
``closed topological vertex''), a Kontsevich-like description via random
matrix ensembles, and physical applications for the geometric engineering of
extended $\cN=2$ $U(1)$ gauge theories \cite{Marshakov:2006ii} in five dimensions. We plan to return on some of this points in the near future. \\



{\noindent \bf Acknowledgements.} It is a pleasure to thank {\it in primis}
Boris Dubrovin for his many insightful comments and suggestions, that
crucially helped us to solve many puzzles in this project. The present work
would not have been undertaken without his influence. I am moreover grateful
to Vincent Bouchard, Motohico Mulase and Brad Safnuk for  inviting me to
present part of this material at the AIM workshop ``Recursion structures in
topological string theory and enumerative geometry'' in Palo Alto; I would
also like to thank the participants for their valuable comments. I am likewise
happy to thank Tom Coates for inviting me to visit Imperial College in
November 2009, as well as for his interest, support, and helpful discussions
on related subjects. I also benefitted from discussions with Guido Carlet,
Renzo Cavalieri and Paolo Rossi. \\
This work was supported by a post-doc fellowship of the Fonds National Suisse (FNS); partial support from a ``Progetto Giovani 2009'' grant of the Gruppo Nazionale per la Fisica Matematica (GNFM) is also acknowledged.

\section{The genus zero theory of local $\bbP^1$ and integrable hierarchies}
\label{sec:displess}
\subsection{Hamiltonian integrable hierarchies from associativity equations}
\label{sec:disphier}
In this section we sketchily review the general construction of dispersionless Hamiltonian hierarchies from associativity equations. The details can be found in the original literature on the subject \cite{Dubrovin:1992dz, Dubrovin:1994hc}; see also \cite{rossi-2009} for a recent and very readable account of this material. \\
Let $\cV$ be a $n$-dimensional vector space over a field $\bbK$. We will denote by $\cN :=\LL(S^1, \cV)=\{\mathbf{u}:S^1 \to\cV\}$ the formal loop space of $\cV$; the components of the formal maps $\mathbf{u} \in \cN$ will often be written as $u^\a(x)$, where $x\in S^1$ and $\a=1,\dots, n$. $\cN$ carries naturally the structure of a linear space over $\bbK$; a distinguished subspace of its dual $\cN^*$ is given by the so called {\it local functionals} 
\beq
F[\bfu] := \int_{S^1} f(x, \bfu,\bfu_x,\bfu_{xx},\ldots,\bfu^{(k)}, \ldots) \rd x,
\eeq
where $\bfu^{(k)}$ denotes the $k^{\rm th}$ $x$-derivative of $\bfu$. The adjective ``local'' refers to the fact that we require the density $f$ to be a {\it differential polynomial}, i.e., to depend polynomially\footnote{In this formal setting and in absence possibly of a well defined analytic theory of functions when $\bbK\neq\bbC$ or $\bbR$, a non-polynomial functional dependence should be thought of as a non-truncating formal power series expansion in $u^\a(x)$.} on $\bfu^{(k)}$ for $k>0$. The set of local functionals on $\cN$ will be called $LF(\cN)$. We want to define a Hamiltonian infinite dimensional dynamical system on $\cN$ via the following data:
\bit
\item a local Poisson bracket 
\beq
 \l\{ u^\a(x),  u^\b(y) \r\} = \sum_{j=0}^m a^{\a\b}_j(\bfu,\bfu_x,\bfu_{xx},\ldots,\bfu^{(n)}, \ldots) \delta^{(j)}(x-y),
\label{eq:parpoisgen}
\eeq
for some integer $m\in \bbN$ and differential polynomials $a^{\a\b}_{j}$; we have denoted by $\delta^{(j)}(x-y)$ the $j^{\rm th}$ distributional derivative of Dirac's $\d$-function. By bilinearity and the functional Leibnitz rule, the Poisson bracket of elements $F, G \in \cN$ is
\beq
\l\{ F, G\r\} = \int_{S^1 \times S^1} \frac{\d F}{\d u^\a(x)} \frac{\d G}{\d u^\b(y)} \l\{ u^\a(x),  u^\b(y) \r\} \rd x \rd y.
\eeq
The Poisson structure on $\cN$ is said to be of {\it hydrodynamic type} if $a^{\a\b}_j= \delta_{j1}\eta^{\a\b}$ for a constant, symmetric, non-degenerate matrix $\eta^{\a\b}$;
\item Hamiltonian flows on $\cN$ generated by Hamiltonians $H[\bfu]\in LF(\cN)$ via
\beq
\bfu_{t} = \{\bfu, H[\bfu]\}.
\eeq
\eit
\begin{defn}
Let $ \l\{ u^\a(x),  u^\b(y) \r\}$ be a hydrodynamic Poisson bracket on $\cN$
and $H_{\a,p}[\bfu] \in LF(\cN)$, $\a=1,\dots,n$, $p \in \bbN$, be a countably infinite sequence of independent local functionals
\beq
H_{\a,p}[\bfu]=\int_{S^1}h_{\a,p}(\bfu(x), \bfu_x(x), \ldots,x)\rd x.
\eeq
Then:\\
\ben
\item the equations 
\beq
\frac{\de \bfu}{\de t^{\a,p}} = \{\bfu, H_{\a,p}[\bfu]\}
\label{eq:hierarchygen}
\eeq
are said to make up a {\it Hamiltonian integrable hierarchy of PDEs} if they satisfy the involutivity condition
\beq
\{H_{\a,p}[\bfu], H_{\b,q}[\bfu]\}=0 \quad \forall \a,\b,p,q;
\eeq
\item the hierarchy is said to possess a $\tau$-structure it there exists a potential $\de_x\ln \tau$ for the integrability condition
\beq
\de_{t^{b,q}} h_{\a,p-1}=\de_{t^{\a,p}}
h_{\b,q-1}=\de_x\de_{t^{\a,p}}\de_{t^{b,q}}  \ln \tau.
\eeq
$\tau(\bfu,\bfu_x,\bfu_{xx},\ldots,\bfu^{(n)}, \ldots)$ is called a $\tau${\it-function} of the hierarchy;
\item the hierarchy is  said to be {\it dispersionless} if the system \eqref{eq:hierarchygen} is quasi-linear, i.e., if the densities
$h_{\a,p}$ do not depend on derivatives $\bfu^{(k)}$ of the fields for $k\geq 1$.
\een
\end{defn}
It was suggested by Witten \cite{Witten:1990hr} that Conjecture
\ref{conj:integrableGW} should have a description in this framework, with the
Hamiltonian densities $h_{\a,p}$ being related to 2-point ``big phase space''
correlators in a topological field theory coupled to gravity. For the genus
zero theory this was formalized in fairly large generality, and in a
completely explicit way, in the work of Dubrovin \cite{Dubrovin:1992dz,
  Dubrovin:1994hc}. We will now review it in the case we will be interested in
of the $T\simeq (\bbC^*)^k$-equivariant Gromov--Witten theory of a K\"ahler target manifold
$X$. We will assume that $T$ acts with compact fixed loci $F$. \\
Take $\cV_X:=QH_T^\bullet(X)$ to be the big equivariant quantum co-homology ring of $X$; in this case $\bbK = \bbC(\lambda_1, \dots, \lambda_k)$ is the field of fractions of $H_T^\bullet(\mathrm{pt})$. Suppose moreover that $\cV_X^{\mathrm{odd}}=0$, and pick a basis $\phi_\a$, $\a=1,\dots, h_X$ of $\cV_X$, where $h_X=\dim_{\bbK} \cV_X$ and $\phi_1 = \mathbf{1}_{\cV_X}$; a generic element of $\cV_X$ will be written $\bfu=\sum_\a u^\a \phi_\a$ with $u^\a\in \mathbb{K}$. The genus zero primary Gromov--Witten potential of $X$ is a formal analytic function $F_0:\cV_X\to \mathbb{K}$,
\beq
F_0(\bfu) = \sum_{j,d=0}^\infty \frac{1}{j!}\Big\langle \overbrace{\bfu, \dots, \bfu}^{j \hbox{ \footnotesize times}} \Big\rangle_{0,j,d}^X = \sum_{j=0}^\infty \sum_{\substack{\a_1 \dots \a_n}} f_{\a_1 \dots \a_j} u^{\a_1}\dots u^{\a_j},
\label{F0series}
\eeq
satisfying:
\ben
\item $\eta_{\a\b}:=\partial_1\partial^2_{\a\b}F_0(\bfu)$ is a nondegenerate, constant symmetric matrix;
\item $F_0$ obeys the following set of third order, non-linear PDEs
\beq
c_{\a\b\g} \eta^{\g\d} c_{\d\e\zeta} =c_{\a\e\g}  \eta^{\g\d} c_{\d\b\zeta},
\label{wdvv}
\eeq
where the three-point function $c_{\a\b\g}$ is defined as $c_{\a\b\g}:=\partial^3_{\a\b\g} F_0$.
It is intended that indices are raised with the non-degenerate contravariant 2-tensor $\eta^{\a\b}=(\eta^{-1})_{\a\b}$, and we use Einstein's convention to sum over repeated indices.
\een

\begin{rem}
\label{rem:euler}
It should be stressed at this point that, as opposed to the usual definition of a Frobenius manifold \cite{Dubrovin:1994hc}, we do not have a quasi-homogeneity condition obeyed by $F_0$. This is due to the fact that the ground field $\bbK$ has a non-trivial grading in this case, and therefore the natural Euler operator, keeping track of the equivariant de Rham degree, is not $\bbK$-linear.
\end{rem}
\noindent Eq. \eqref{wdvv} implies the following fact. Define the 1-parameter family of connections on $T^*\cV_X$
\beq
D_z := d + \Gamma,
\eeq
where the Christoffel symbol $\Gamma_\a$  in components reads $(\Gamma_\a)_\b^\gamma := z c_{\a\b}^\g$ and $z\in\mathbb{K}$. Notice that because of integrability of $c_{\a\b\g}$ and \eqref{wdvv}  we have
\beq
D_z^2=0 \quad \forall z,
\label{eq:flatcond}
\eeq
that is, the connection is flat. Its horizontal sections $\omega^{(\b)}_\a d u^\a = d h^{(\b)}$, where $(\b)=1,\dots, h_X$ labels a fundamental set of solutions of \eqref{eq:flatcond}, should come from a basis $f^{(\b)} \in \mathrm{Fun}(\cV_X)$ of solutions of the holonomic system of PDEs
\beq
\partial^2_{\alpha \beta} h^{(\d)} = z c_{\alpha \beta}^\gamma \partial_{\gamma} h^{(\d)}, \quad \d=1, \ldots, h_X.
\label{flatfunct}
\eeq
We will call the solutions of \eqref{flatfunct} the {\it flat functions} of $\cV_X$. Their duals $h_\a(u,z):=\eta_{\a\b} h^{\b} (u,z)$ can always be normalized such that
\bea
\label{eq:norm1}
h_\a(\bfu,0)&=&w_\a=\eta_{\a\b}u^\b, \\
\label{eq:norm2}
\de_\g h_\a(\bfu,z)\eta^{\g\d} \de_\d h_\b(\bfu,-z) &=& \eta_{\a\b}, \\
\label{eq:norm3}
\de_1 h_\a(\bfu,z) &=& z h_\a(\bfu,z) + \eta_{1\a}.
\eea
\begin{rem}
  Eqs. \eqref{eq:norm1}-\eqref{eq:norm3} do not fix completely the ambiguity in the choices of the $z$-dependent constants of integration of \eqref{flatfunct}. In the ordinary Frobenius manifold case such ambiguity could be dealt with by imposing additional conditions coming from the existence of the Euler vector field. In the cases we are interested in such a procedure will have to be performed otherwise (see Sec. \ref{sec:conifold}).
\end{rem}
Solutions of WDVV relate to the theory of Hamiltonian dispersionless systems in the following way. Endow the loop space $\cN_X := L(S^1, \cV_X)$ with the hydrodynamic Poisson bracket
\beq
\l\{ u^\a(x),  u^\b(y) \r\} = \eta^{\a\b} \delta'(x-y).
\label{eq:poissbrack}
\eeq
Then the Taylor coefficients of the $z$-expansion of $h_\a(\tau; z)$ with respect to $z$,
\beq
h_\a(\bfu,z) =: \sum_{z=0}^\infty h_{\a,p-1}(\bfu) z^p,
\label{eq:hap}
\eeq 
define dispersionless Hamiltonian densities on $\cN_X$. The system of $1^{\rm st}$ order quasi-linear PDEs 
\beq
\frac{\de \bfu}{ \de t^{\a,p}} = \l\{\bfu, \int_{S^1}h_{\a,p}(\bfu(x))\r\} \rd x
\label{eq:princhier}
\eeq
will be called the {\it Principal Hierarchy} of $X$. We have the following
\begin{thm}[Dubrovin]
\label{thm:dubr}
The set of Hamiltonians $H_{\a,p}=\int_{S^1}h_{\a,p} \rd x $ mutually Poisson--commute with respect to the Poisson bracket \eqref{eq:poissbrack}. Let $u^\alpha(\mathbf{t})$ solve the system (\ref{eq:princhier}) with boundary condition 
\beq
\label{eq:initial}
u_\alpha(\mathbf{t})\bigg|_{\substack{t^{a,p}=0 \\ \hbox{\rm \footnotesize for } p>0}}
=\de^2_{1,\a} F_0(t^{1,0}+x,t^{2,0},t^{3,0},\dots..., t^{n,0})
\eeq
and define for all times
\bea
\bra\bra\tau_{p}(\phi_\a) \tau_q (\phi_{\b}) \ket\ket_0 &:=& \frac{1}{2\pi i} \oint
\oint \frac{dzdw}{z^{p+1}w^{q+1}} \frac{ \de_\g h_\a(\bfu(\mathbf{t}),z)\eta^{\g\d} \de_\d h_\b(\bfu(\mathbf{t}),w)- \eta_{\a\b}}{z+w}, \nn \\ \\
\mathcal{F}_0(x+t^{1,0},t^{2,0}, \dots ) &:=& \frac{1}{2} \sum\bra\bra\tau_{p}(\phi_\a) \tau_q (\phi_{\b}) \ket\ket_0(\mathbf{t}) t^{\a,p} t^{\b,q} + \sum \bra\bra\tau_{p}(\phi_\a) \tau_1 (\phi_1) \ket\ket_0(\mathbf{t}) t^{\a,p} \nn \\ &+& \frac{1}{2} \bra\bra\tau_{1}(\phi_1) \tau_1 (\phi_{1}) \ket\ket_0(\mathbf{t}), \\
\langle\langle \phi_{\alpha,p} \phi_{\beta,q}\dots \rangle\rangle_0 &:=& \partial_{t^{\alpha,p}} \partial_{t^{\beta,q}}\dots\partial_{\dots} \mathcal{F}_0(\mathbf{t}).
\label{eq:dercorr}
\eea
Then $\mathcal{F}_0$ is the logarithm of a $\tau$ function for the hierarchy \eqref{eq:princhier}. It moreover satisfies
\beq
\begin{array}{cccr}
\mathcal{F}_0|_{t^{\a,p}=0 \hbox{ \rm \scriptsize for } p>0} &=& F_0(t^{\a,0}) & \hbox{\rm (reduction to primaries)} \\
\partial_x \mathcal{F}_0 &=& \sum t^{\alpha,p}\partial_{t^{\alpha,p-1}} \mathcal{F}_0 +\frac{1}{2} \eta_{\alpha \beta} t^{\alpha,0} t^{\beta,0} & \hbox{\rm (string equation)}\\
\langle\langle\phi_{\alpha,p} \phi_{\beta,q} \phi_{\gamma,r} \rangle\rangle_0 &=&\langle\langle \phi_{\alpha,p-1} \phi_{\delta,0}\rangle\rangle_0\eta^{\delta \epsilon} \langle\langle \phi_{\epsilon,0} \phi_{\beta,q} \phi_{\gamma,r} \rangle\rangle_0 & \hbox{\rm (genus zero TRRs)}
\end{array}
\label{eq:dubrTRR}
\eeq
\end{thm}
For the purpose of the Gromov--Witten/Integrable Systems correspondence this construction has a number of very attractive features, together with a few weak points. The main virtue of this construction is that it does not depend on the details of $X$, apart from the requirement that $\cV_X^{\mathrm{odd}}=0$; 
moreover, it provides an explicit construction of the integrable hierarchy starting from primary data, thus yielding a constructive proof of Conjecture \ref{conj:integrableGW} at the leading order in $\e$ (i.e. in the genus zero subsector). However, to make it work we need to have control on $F_0$ in {\it closed} form - no implicit, recursive or up-to-inversion-of-the-mirror-map form will do the job. Any polynomial truncation of (\ref{F0series}) affects dramatically the form of the three-point couplings $c_{\a\b\g}$ and therefore the flat functions. In other words, we must know explicitly {\it all} the coefficients $f_{\a_1\dots \a_j}$ in (\ref{F0series}). This limitation turns out to be very constraining in the context of ordinary Gromov--Witten theory, where it basically reduces the list of viable examples to the cases of $X=\mathrm{pt}$ and $X=\bbP^1$ we mentioned in Sec. \ref{sec:introgen}. However, since the construction does not depend on the existence of an Euler vector field, we might expect to find new examples in the context of equivariant Gromov--Witten theory. Indeed, as we are going to argue, the local theory of rational curves inside Calabi--Yau threefolds evades this limitation in a large number of cases.

\subsection{The resolved conifold}
\label{sec:dispconif}
Let us then consider the target spaces $X_k$ of \eqref{eq:Xk}. We begin with
the rigid case $k=1$, and consider its equivariant theory with respect to a
$T\simeq (\bbC^*)^2$ fiberwise action rescaling the fibers. Let $H(X_1) :=
H_T^\bullet(X_1) \simeq  H^\bullet(F_1) \otimes \mathbb{C}(\lambda_1,
\lambda_2)$ denote the localized $T$-equivariant cohomology of $X_1$ and $F_1
\simeq \mathbb{P}^1$ be the fixed locus of the $T$-action, that is, the null
section of $X_1\to \mathbb{P}^1$. Let moreover $(1,p)$ denote the canonical
basis of $H(X_1)$ (regarded as a free $\mathbb{C}(\lambda_1,
\lambda_2)$-module), where $1$ and $p$ denote respectively the lifts to
$T$-equivariant co-homology of the identity and the K\"ahler class of the base $\bbP^1$, and write $\bfu =: v+w p$, i.e. $v:= u^1$, $w:=u^2$ with $v,w \in \mathbb{C}(\lambda_1,\lambda_2)$. We separate the degree zero (``classical'') and positive degree (``quantum'') parts of the genus zero Gromov--Witten potential of $X_1$ as
\beq
F^{X_1}_0(\bfu) =  F^{X_1}_{0, \mathrm{cl}}(\bfu) + F^{X_1}_{0,\mathrm{qu}}(\bfu),
\label{treesplit}
\eeq
where
\bea
 F^{X_1}_{0, \mathrm{cl}}(\bfu) &=&  \frac{1}{3!}\int_{[\mathbb{P}^1]} \frac{\bfu \cup \bfu \cup \bfu}{e(\mathcal{N}_{X_1/F_1})}, \nn \\
 F^{X_1}_{0,\mathrm{qu}}(\bfu) &=& \sum_{d >0} e^{d w} N^{(1)}_{0,d}, \nn \\
N^{(1)}_{g,d} &=& \int_{[(X_1)_{g,0,d}]^{vir}} 1. 
\eea
A special feature of the $(\bbC^*)^2$-equivariant theory of the resolved conifold is that the invariants $N^{(1)}_{0,d}$ have a closed expression for all $d$ \cite{MR2276766}
given by the Aspinwall-Morrison multi-covering formula \cite{Aspinwall:1991ce}
\beq
N^{(1)}_{0,d}=\frac{1}{d^3}.
\label{eq:aspmorr}
\eeq
$X_1$ then belongs to the list of fortunate cases where a closed form expression for the genus zero Gromov--Witten invariants of all degrees, and therefore for the prepotential, is known in terms of special functions. Explicitly we have
\bea
F^{X_1}_{0, \mathrm{cl}}(v,w) &=& \frac{1}{3!}\int_{[\mathbb{P}^1]} \frac{(v+w p)^3}{(\lambda_1-p)(\lambda_2-p)} \nn \\ &=& \frac{1}{3! \lambda_1 \lambda_2}\int_{[\mathbb{P}^1]} (v+w p)^3\l(1+\l(\frac{1}{\lambda_1}+\frac{1}{\lambda_2}\r)p\r) 
\label{clterm1}
\eea
and hence
\beq
F_0^{X_1}= \frac{v^3}{3!}\frac{\lambda_1+\lambda_2}{\lambda_1^2 \lambda_2^2} +\frac{1}{2 \lambda_1 \lambda_2} v^2 w +\Li_3(e^w),
\label{eq:prepconifold}
\eeq
where we have introduced the polylogarithm function
\beq
\Li_j(x) = \sum_{n=1}^\infty \frac{x^n}{n^j}.
\eeq
This is all is necessary to apply the machinery of Sec. \ref{sec:disphier}. For future use, we state the following
\begin{lem} Consider the following solution of WDVV
\beq
F_0= \frac{Pv^3}{3!} +\frac{Qv^2 w}{2}  +\Li_3(e^w).
\eeq
Then the general integral of the flatness conditions \eqref{flatfunct} reads
\beq 
f(v,w,z,P,Q) = A(w,z,P,Q)\frac{e^{v z}}{z}+ B(z),
\label{eq:genintflat}
\eeq
where
\bea
A(w,z,P,Q) &=&
c_1(z) \, _2F_1\left(-\Delta_+,-\Delta_-;1;e^w\right) \nn \\ &+& c_2(z) \, _2F_1\left(1+\Delta_-,1+\Delta_+;\frac{z P  }{Q ^2}+2;1-e^w\right)
   \left(1-e^w\right)^{\frac{z P  }{Q ^2}+1},
\label{eq:genintA} \\
\Delta_\pm &=&\frac{z \left(P \pm\sqrt{P^2-4 Q
   ^3}\right) }{2 Q ^2}. \nn
\eea
\end{lem}
\pf The form \eqref{eq:genintflat} follows from \eqref{flatfunct} with $\alpha=v$. Putting $\a=\b=w$ in \eqref{flatfunct} yields a Fuchsian ODE for $A$ as a function of $w$, 
\beq
\de^2_{ww}A =\frac{e^w }{1-e^w}\left(\frac{z^2 A(w)}{Q}-\frac{P z \de_w A(w)}{Q^2}\right),
\eeq
whose general integral has the form \eqref{eq:genintA}.\begin{flushright}$\square$\end{flushright}
For the prepotential \eqref{eq:prepconifold} we have 
\beq
\Delta_+=z \lambda_1, \quad \Delta_-=z\lambda_2, \quad
\frac{P}{Q^2}=\lambda_1+\lambda_2.
\eeq
Let us fix a normalization of the corresponding flat functions $h^\a(v,w;z)$,
\beq
h^\a(v,w;z)  = A^\alpha(w,z,\lambda_1, \lambda_2)\frac{e^{v z}}{z}+ B^\alpha(z),
\label{eq:unnormff}
\eeq
in order for the flows to satisfy the string axiom and the genus zero TRRs. Eq. \eqref{eq:norm3} fixes $B^\alpha(z)$ to be
\beq
B^\a(z)=-\frac{\delta^{v,\a}}{z}.
\eeq
To fix completely $A^\alpha(w,z,\a,\b, \lambda)$ we use the fact that it is related \cite{MR1677117} to the fundamental solution $S_{\a\b}$ of the Gauss-Manin system as
\beq
A^\alpha=\de_v h^\a\Big|_{v=0},
\eeq
\beq
\de_v h^\a=:S_0^\a=:J^\a,
\eeq
that is to say, it corresponds to the $\a$-component of the $J$-function at $v=0$. The Coates--Givental theorem \cite{MR2276766} prescribes it to take the form
\beq
\label{eq:Jfunconifold}
J(v=0,w, z, \lambda_1, \lambda_2)=e^{z p \log{q(w)}} \sum_{d\geq 0}\frac{\prod_{m=-d+1}^0\left(-p+m/z+\lambda_1\right)\left(-p+m/z+\lambda_2\right)}{\prod_{m=1}^d\left(p+m/z\right)^2}q(w)^d,
\eeq
where $q(w)$ is the inverse mirror map. We have the following
\begin{prop} 
\label{prop:norm}
For the normalized flat functions \eqref{eq:unnormff} we have
\bea
A^\alpha(w,z,\lambda_1, \lambda_2) &=& c^\alpha_1(z,\lambda_1, \lambda_2) \, _2F_1\left(-z \text{$\lambda_1$},-z \text{$\lambda_2$};1;e^w\right) \nn \\ &+& c^\a_2(z, \lambda_1,\lambda_2) \, _2F_1\left(z \text{$\lambda_1$}+1,z \text{$\lambda_2$}+1;z (\text{$\lambda_1$}+\text{$\lambda_2$})+2;1-e^w\right) \nn \\ & \times &  \left(1-e^w\right)^{z (\text{$\lambda_1$}+\text{$\lambda_2$})+1},
\label{eq:Anorm}
\eea
where:
\bea
\label{eq:cv1}
c^v_1(z,\lambda_1, \lambda_2) &=& 1, \\
c^v_2(z,\lambda_1, \lambda_2) &=& 0, \\
\label{eq:cw1}
c^w_1(z,\lambda_1, \lambda_2) &=& -z \l[\psi ^{(0)} (z \lambda_1+1)+\psi ^{(0)}(z \lambda_2+1)+2 \gamma\r],   \\
c^w_2(z,\lambda_1, \lambda_2) &=&  -\frac{z \Gamma (z \lambda_1+1) \Gamma (z \lambda_2+1)}{\Gamma (z (\lambda_1+\lambda_2)+2)}.
\label{eq:cw2}
\eea
In \eqref{eq:cw1}, $\gamma$ is the Euler-Mascheroni constant, while $\psi ^{(0)}(x)$ is the polygamma function 
\beq
\psi ^{(0)}(z)=\frac{d\log{\Gamma(z)}}{dz}
\eeq
\label{thm:normflat}
\end{prop}
\noindent {\it Proof.} The $\mathcal{O}(z)$ term of the expansion of the $J$-function \eqref{eq:Jfunconifold} is the statement that the mirror map is {\it trivial} in this case
\beq
\log{q}=w \hbox{ (mod } 2\pi i).
\eeq
Let us examine the summand in \eqref{eq:Jfunconifold} above more closely, starting from the numerator. The finite product gives, remembering that $p^2=0$,
\beq
\bary{l}
\prod_{m=-d+1}^0\left(-p+m/z+\lambda_1\right)\left(-p+m/z+\lambda_2\right) =  \\
\prod_{m=0}^{d-1}\l[p \left(\frac{2 m}{z}-\lambda_{1}-\lambda_{2}\right)+\frac{m^2-z \lambda_{1} m-z \lambda_{2} m+z^2 \lambda_{1} \lambda_{2}}{z^2}\r]=
\frac{\left(\frac{1}{z^2}\right)^d \Gamma (d-z \lambda_{1}) \Gamma (d-z \lambda_{2})}{\Gamma (-z \lambda_{1}) \Gamma (-z \lambda_{2})} \\
-\l[\frac{\left(\frac{1}{z^2}\right)^d z \Gamma (d-z \lambda_{1}) \Gamma (d-z \lambda_{2}) \left(\psi ^{(0)}(-z \lambda_{1})-\psi ^{(0)}(d-z
   \lambda_{1})+\psi ^{(0)}(-z \lambda_{2})-\psi ^{(0)}(d-z \lambda_{2})\right)}{\Gamma (-z \lambda_{1}) \Gamma (-z \lambda_{2})}\r]p, 
\eary
\eeq
while for the inverse of the denominator we obtain simply
\beq
\frac{1}{\prod_{m=1}^d\left(p+m/z\right)^2}= \frac{\left(z^2\right)^d}{\Gamma
  (d+1)^2}-\frac{2 z \left(z^2\right)^d H_d}{\Gamma (d+1)^2}p,
\eeq
where $H_d$ is the $d^{\rm th}$ harmonic number. For the $v$-component, this means that we should have
\beq
A^v(w,z,\lambda_1,\lambda_2)=
\sum_{d\geq 0} \frac{e^{wd} \Gamma (d-z \lambda_{1}) \Gamma (d-z \lambda_{2})}{\Gamma (d+1)^2 \Gamma (-z \lambda_{1}) \Gamma (-z \lambda_{2})}.
\eeq
Comparison with \eqref{eq:Anorm} sets
\beq
c^{v}_1 =1, \qquad c^{v}_2=0.
\eeq
On the other hand, the component $J^{w}$ of the $J$-function in the direction of the volume form is a series that looks as follows
\bea
J^{w}(0, w,z,\lambda_1,\lambda_2) &=& z w A^v(w,z,\lambda_1,\lambda_2) + 
\sum_{d\geq 0} \Bigg[ e^{w d}\frac{z \Gamma (d-z \lambda_{1}) \Gamma (d-z \lambda_{2})}{\Gamma (d+1)^2 \Gamma (-z \lambda_{1}) \Gamma (-z \lambda_{2})} \nn \\ &\times&  \left(-2 H_d + \sum_{i=1,2}\l(\psi ^{(0)}(d-z \lambda_{i}) - \psi ^{(0)}(-z \lambda_{i})\r)\right) \Bigg].
\eea
The term proportional to $z w$  term comes from the $e^{zp\log{q}}$ prefactor of the $I$ function. Let us then fix the coefficients $c^{w}_i(z)$ by Taylor-expanding \eqref{eq:Anorm} at $q=\exp{w}=0$. We get an expansion of the form $a \log{q}+b+o(1)$
\bea
A^w(w,z,\lambda_1,\lambda_2) &=& 
- \frac{c^w_2(z,\lambda_1,\lambda_2) \Gamma (z
    (\lambda_{1}+\lambda_{2})+2)}{\Gamma (z \lambda_{1}+1) \Gamma (z
    \lambda_{2}+1)} \log (q)  -c^w_1(z,\lambda_1,\lambda_2) \nn \\ &-& 
\frac{c^w_2(z,\lambda_1,\lambda_2) \Gamma(2+ z \lambda_1 + z \lambda_2)(
  \psi^{(0)}(z \lambda_{1}+1)+\psi^{(0)}(z \lambda_{2}+1)+2 \gamma)}{\Gamma (z
  \lambda_{1}+1) \Gamma (z \lambda_{2}+1)} \nn \\ &+& \mathcal{O}(q),
\eea
while from the explicit form of the $J$-function we get
\beq
A^w(w,z, \lambda_1, \lambda_2)=z \log{q} + \mathcal{O}(q).
\eeq
Matching the logarithmic coefficient gives \eqref{eq:cw2}, while the $\mathcal{O}(1)$ term yields \eqref{eq:cw1}. This completely fixes the form of the deformed flat coordinates; it is straightforward to check that the normalization conditions \eqref{eq:norm1}-\eqref{eq:norm3} are satisfied. \begin{flushright}$\square$\end{flushright}
Theorem \ref{thm:dubr} and Proposition \ref{thm:normflat} together complete the construction of the dispersionless hierarchy that governs the genus zero Gromov--Witten theory of the resolved conifold. To see what its first flows look like, take the $z$-expansion of the densities (see \eqref{E-expansion:1}-\eqref{E-expansion:2})
\bea
h_v(v,w,z,\lambda_1,\lambda_2) &=&
\frac{w \lambda_{1} \lambda_{2}+v (\lambda_{1}+\lambda_{2})}{\lambda_{1}^2 \lambda_{2}^2}+\frac{1}{\lambda_{1}^2 \lambda_{2}^2}\Bigg[\frac{1}{2}
   (\lambda_{1}+\lambda_{2}) \left(v^2+2 \lambda_{1} \lambda_{2} \text{Li}_2\left(e^w\right)\right) \nn \\ &-& \frac{1}{6} \lambda_{1}
   \lambda_{2} \left(-6 v w-6
     (\lambda_{1}+\lambda_{2}) \l(\text{Li}_2\left(1-e^w\right)+w \Li_1(w)+\frac{\pi^2}{6}\r)\right)\Bigg] z \nn \\ &+& O\left(z^2\right), \\
h_w(v,w,z,\lambda_1,\lambda_2) &=&
\frac{v}{\lambda_{1} \lambda_{2}}+\left(\frac{v^2}{2 \lambda_{1} \lambda_{2}}+\text{Li}_2\left(e^w\right)\right) z+O\left(z^2\right).
\eea
The first two flows are then
\bea
\frac{\de v}{\de t_{1,0}} &=& v_x, \\
\frac{\de w}{\de t_{1,0}} &=& w_x, \\
\frac{\de v}{\de t_{2,0}} &=& \lambda_1 \lambda_2 \frac{e^w}{1-e^w} w_x, \\
\frac{\de w}{\de t_{2,0}} &=&  v_x -(\lambda_1+\lambda_2) \frac{e^w}{1-e^w} w_x.
\label{eq:firstflowsconifold}
\eea
Eliminating $v$ and putting $t:=t_{2,0}$ we obtain the non-linear wave equation
\beq
w_{tt}=\lambda_1 \lambda_2 \l(\frac{e^w}{1-e^w} w_x\r)_x-(\lambda_1+\lambda_2) \l(\frac{e^w}{1-e^w} w_x\r)_t.
\label{eq:nlinconifold}
\eeq
In Sec. \ref{sec:ablo} we will see how this relates, in one notable case, to known examples in the theory of integrable hierarchies.
\label{sec:conifold}
\subsection{The diagonal action}
\label{sec:diagonal}
Let us move on to the general case  \eqref{eq:Xk} of $X_k$. In the first place we restrict $T$ to be isomorphic to a one dimensional torus acting diagonally on the two fibers. We adapt, with obvious meaning of symbols, the conventions of Sec. \ref{sec:conifold} for the potentials, the genus zero invariants, and the equivariant co-homology classes of $X_k$, appending an index $k$ and a superscript $di$ (for ``diagonal'') whenever necessary. \\

The choice of a diagonal action is special for two reasons. First, this is the case that corresponds to the invariants defined by Bryan and Pandharipande in \cite{Bryan:2004iq}. Secondarily, it surprinsigly turns out to be a {\it subcase} of the $(\bbC^*)^2$-equivariant theory of the resolved conifold we treated in the previous section. The quantum tail of the prepotential indeed \cite{MR2276766, Forbes:2006sj} has for all $k$ the Aspinwall-Morrison like form
\beq
N^{(k, {\rm di})}_{0,d}=\frac{1}{d^3}.
\eeq
On the other hand, the classical piece is given by
\bea
F^{X_k, {\rm di}}_{0, \mathrm{cl}}(v,w) &=& \frac{1}{3!}\int_{[\mathbb{P}^1]} \frac{(v+w p)^3}{(\lambda-kp)(\lambda+kp-2p)} \nn \\ &=& \frac{1}{3! \lambda^2}\int_{[\mathbb{P}^1]} (v+w p)^3\l(1+\frac{2p}{\lambda}\r) 
\label{cltermk}
\eea
and hence
\beq
F_0^{X_k, {\rm di}}= \frac{1}{3}\l(\frac{v}{\lambda}\r)^3 +\frac{1}{2 \lambda^2} v^2 w +\Li_3(e^w) \qquad \forall k \in \bbZ.
\label{eq:prepxkdi}
\eeq
Therefore, our results in the previous section apply, {\it a fortiori}, to the theory with generic $k$ and $\lambda_1=\lambda_2$.

\subsection{The anti-diagonal action}
Another case of special interest is given by the reduction to the case of a $T\simeq\bbC^*$ fiberwise action with {\it opposite} characters on the two fibers. In this case, the equivariant Euler class of $X_k$ is trivial
\beq
e_T(X_k)=0,
\eeq
that is, $X_k$ is equivariantly Calabi--Yau. The notation will follow the same conventions as in the previous two sections, with a superscript $ad$ for ``anti-diagonal'' added whenever needed. \\

It was conjectured in general for toric Calabi--Yau threefolds, and verified explicitly for the case at hand \cite{Bryan:2004iq}, that the invariants in the equivariantly Calabi--Yau case are the ones that most closely make contact with the physics prediction based on topological open/closed duality. In particular the authors of \cite{Bryan:2004iq} could prove the following formula, which could be regarded as a specialization to $X_k$ of the topological vertex formalism of \cite{Aganagic:2003db}:
\begin{thm}[Bryan-Pandharipande]
\label{t:bryanp}
The {\rm fixed-degree $d>0$, all-genus} Gromov - Witten potentials of $X_k^{\rm ad}$ are given by the following sum over partitions
\beq
\label{bryanp}
\sum_{g\geq 0} \epsilon^{2g-2} N^{(k), {\rm ad}}_{g,d} = (-1)^{d(k-1)}\sum_\rho \left(\frac{\mathrm{dim}_Q \rho}{d!}\right)^2 Q^{c_\rho(1-k)}.
\eeq
In (\ref{bryanp}), $\rho$ is a Young diagram (a partition of length
$l(\rho)$), $c_\rho$ is its total content, $Q:=e^{i\epsilon}$, $h(\Box)$ is
the hooklength of a box in $\rho$ and 
\beq
\frac{\mathrm{dim}_Q \rho}{d!} = \prod_{\Box \in \rho} \left(2
\sin{\frac{h(\Box)\epsilon}{2}} \right)^{-1}.
\eeq
\end{thm}

\noindent As we stressed in Sections \ref{sec:introresults} and \ref{sec:disphier}, a key point in our analysis is the construction of the  hierarchy governing the genus zero theory starting from a {\it closed-form} solution of WDVV. To see this, we should be able to obtain a closed expression for the {\it all-degree, genus zero} invariants starting from \eqref{bryanp}. This is the content of the next
\begin{prop}[\cite{Caporaso:2006gk}]
The quantum part  $F^{X_k,{\rm ad}}_{0,qu}(w)$ of the $A$-model prepotential of $X_k$ with anti-diagonal action is 
\bea
\label{potX1}
F^{X_1,{\rm ad}}_{0, qu}(w) &=& \Li_3(e^{w}),  \\
\label{potX2}
F^{X_2,{\rm ad}}_{0, qu}(w) &=& -\Li_3(e^{w}),  \\
\label{potXk}
F^{X_k,{\rm ad}}_{0, qu}(w) &=& (-)^{k-1}e^{-w} {}_{n_k+3}F_{n_k+2}\Bigg[1,1,1,1,\frac{1}{n_k},\frac{2}{n_k}, \dots, 1-\frac{1}{n_k};  2,2,2,2, \frac{1}{n_k-1},\nonumber \\ & & \ldots, 1-\frac{1}{n_k-1}; (-1)^k \left(\frac{n_k}{n_k-1}\right)^{n_k-1} n \exp({w})\Bigg]  \qquad (k>2). \nonumber \\
\eea
where $n_k=(k-1)^2$.
\end{prop}
\noindent Eq. \eqref{potX1} is a corollary of \eqref{eq:aspmorr}; the other
two expressions were obtained in \cite{Caporaso:2006gk} by an asymptotic
analysis in $\e$ of the sum over partitions \cite{Bryan:2004iq} based on
matrix model inspired saddle-point techniques applied to the topological
vertex formulae for $X_k$ \cite{Aganagic:2003db}. A mirror symmetry confirmation, based on Birkhoff factorization applied to the Coates--Givental twisted $I$-function, was given in \cite{Forbes:2006ab}. \\
To complete the computation of the prepotential we just have to add the degree zero contribution. We get 
\beq
F_0^{X_k,{\rm ad}}(v, w) = \frac{2-2k}{3!}\left(\frac{v}{\lambda}\right)^3-\frac{1}{2} w \left(\frac{v}{\lambda}\right)^2+F^{X_k,{\rm ad}}_{0, qu}(w).
\label{eq:pottotXk}
\eeq
The case $k=1$ is obviously a reduction of the case of Sec. \ref{sec:conifold} for $\lambda_1=-\lambda_2=\lambda$; inspection shows moreover that the case $k=2$ coincides with the one of Sec. \ref{sec:diagonal} upon sending $F_0 \to -F_0$. \\
The situation for $k>1$ case is instead radically different. A closed form solution for the flat functions seems too hard to obtain; still the Hamiltonian densities $h_{\a,p}$ can be computed and normalized, as we have done before, order by order in $p$. The kind of equations that we find seem totally new: defining the ``Yukawa coupling'' $Y_k(w):= \de^3_{www} F^{X_k,{\rm ad}}_0(w)$ we have
\beq
Y_k(w) =\frac{1}{n_k} -\frac{1}{n_k} {}_{n_k-1}F_{n_k-2}\Bigg[\frac{1}{n_k}, \dots, \frac{n_k-1}{n_k};  \frac{1}{n_k-1},  \ldots, \frac{n_k-2}{n_k-1}; \frac{ (-1)^k n_k^{n_k} e^{-w}}{(n_k-1)^{n_k-1}}\Bigg]
\label{eq:yukk}
\eeq
and we find for instance for the $t:=t_{2,0}$-flow
\bea
\partial_tv(x,t) &=& \left\{v(x,t), H_{w,0} \right\} = Y_k (w) w_x, \\
\partial_tw(x,t) &=& \left\{w(x,t), H_{w,0} \right\}= (v)_x +  (2k-2)Y_k(w) w_x,
\eea
which reduces to a wave equation with hypergeometric\footnote{In fact it was shown in \cite{phdthesis-brini} how to give for $k=3$ a purely algebraic expression for the Yukawa \eqref{eq:yukk}; the final result though sheds little more light on the nature of the equation \eqref{firstflowk}.} non-linearity
\beq
(w)_{tt}=\left(Y_k(w) w_x\right)_x+(2k-2)\left(Y_k(w) w_x\right)_t.
\label{firstflowk}
\eeq

\section{The resolved conifold at higher genus and the Ablowitz--Ladik hierarchy}
In this section we address the problem of deforming the hierarchies we constructed in Sec.~\ref{sec:displess} in order to incorporate higher genus corrections, and we will succesfully find a way to do it in the case $k=1$ with anti-diagonal action. After reviewing in Sec. \ref{sec:disppert} the general problem of constructing Hamiltonian integrable perturbations of dispersionless systems, we will exploit the connection of the Principal Hierarchy of $X_1^{\rm ad}$ with a known integrable lattice to construct a candidate dispersive deformation whose $\tau$-function corresponds to higher genus Gromov--Witten potentials. A quasi-triviality property at $\cO(\e^2)$ will be established in Sec. \ref{sec:quasitr}, and a $\tau$-structure will be defined at this order and used to prove Theorem \ref{thm:main}. Finally in Sec. \ref{sec:g2} we point out the difficulties and subtleties of the higher genus case, and provide a non-trivial $g=2$ test of Conjecture \ref{conj:AL}.
\subsection{Dispersive perturbations of Hamiltonian systems}
\label{sec:disppert}
In the terminology of Sec.~\ref{sec:introresults}, we have performed Step (1) of the construction of the hierarchies relevant to establish Conjecture \ref{conj:integrableGW} for the local theory of $\bbP^1$, at least for the case $k=1$ and for the diagonal and anti-diagonal action. A full answer needs a prescription to perform Step~(2), that is to say, to find a way to unambiguously determine the coefficients $A_{\a,p}^{[g]}$ in \eqref{eq:dispexp}, or within the framework of Hamiltonian hierarchies, the dispersive corrections $H^{[g]}$ of the $g=0$ Hamiltonians
\beq
\label{eq:dispdef1}
H^{\rm disp}_{\a,p}[\bfu, \e]=H^{[0]}_{\a,p}[\bfu] + \e H_{\a,p}^{[1]}[\bfu] + \e^2 H_{\a,p}^{[2]}[\bfu]+\dots,
\eeq
for local functionals $H_{\a,p}^{[n]}[\bfu]$
\beq
H_{\a,p}^{[j]}[\bfu] = \int_{S^1} h_{\a,p}^{[j]}(\bfu, \bfu_x, \bfu_{xx}, \dots, \bfu^{(j)}) \rd x,
\eeq
where $h_{\a,p}^{[j]}$ is a differential polynomial, homogeneous of degree $j$ with respect to the grading \eqref{eq:grading}, and we have appended a superscript $[0]$ to the dispersionless Hamiltonians of the Principal Hierarchy \eqref{eq:princhier}. The statement of integrability is then that
\beq
\l\{H^{\rm disp}_{\a,p}[\bfu, \e], H^{\rm disp}_{\b,q}[\bfu, \e]\r\}=0
\label{eq:dispinv}
\eeq
as a formal power series in $\e$. \\

As we emphasized in Sec. \ref{sec:introresults}, in the context of the equivariant Gromov--Witten theory there are no general methods available to date to determine recursively $H_{\a,p}^{[n]}[\bfu]$ starting from the Hamiltonians of the Principal Hierarchy. However, suppose that a dispersive completion
\beq
\label{eq:dispdef2}
H^{\rm disp}_{\overline{\a},\overline{p}}[\bfu, \e]=\sum_{k=0}^\infty
\epsilon^k H^{[k]}_{\overline{\a},\overline{p}}[\bfu]
\eeq
of {\it one} Hamiltonian $\a=\overline{\a}$, $p=\overline{p}$ be known. We have the following
\begin{thm}[\cite{MR2462355}]
Let $H^{[0]}_{\a,p}[\bfu]$, $\a=1,\dots, n$, $p \in \bbN$ be Hamiltonian local
functionals of a hydrodynamic dispersionless hierarchy of integrable PDEs
\beq
\{H^{[0]}_{\a,p}[\bfu],H^{[0]}_{\b,q}[\bfu]\}=0
\eeq
and let $H^{\rm disp}_{\overline{\a},\overline{p}}[\bfu, \e]$ be a dispersive deformation \eqref{eq:dispdef2} of the Hamiltonian flow $\a=\overline{\a}$, $p=\overline{p}$ for one pair $(\overline{a}, \overline{p})$ and given local functionals $H^{[k]}_{\overline{\a},\overline{p}}[\bfu]$. Then if a dispersive completion of $H^{\rm disp}_{\a,p}[\bfu, \e]$ preserving involutivity of the flows $\forall \e$ exists
\beq \{H^{\rm disp}_{\a,p}[\bfu, \e],H^{\rm disp}_{\b,q}[\bfu, \e]\}=0  \quad \forall \a,\b, p, q, \eeq
it is unique. In such a case, there exists a formal sum of linear differential operators
\bea
\label{doper}
&&
D=\sum_{k=0}^\infty \e^n D^{[k]},
\nn\\
&&
\nn\\
&&
D^{[0]}={\rm id}, \quad D^{[k]} = \sum b^{[k]}_{i_1, \dots, i_n} (u_{1}, \dots, u^{(k)}_{1}, \dots, u_{n}, \dots, u^{(k)}_{n}) 
\frac{\de^{\sum_j i_j}}{\de u_1^{i_1} \dots \de u_n^{i_{n}}},
\eea
such that
\beq
\int_{S^1}D^{[k]} h^{[0]}_{\a,p}(\bfu)\rd x= \int_{S^1}h^{[k]}_{\a,p}(\bfu, \bfu_x, \bfu_{xx}, \dots, \bfu^{(k)}) \rd x = H^{[k]}_{\a,p}[\bfu]
\eeq
satisfies the involutivity condition \eqref{eq:dispinv}. 
In \eqref{doper}, the coefficients $b^{[k]}_{i_1, \dots, i_n}$ are differential polynomials and $\sum_{j=1}^n i_j \leq \left[ \frac{3k}2\right]$.
\label{thm:dop}
\end{thm}
The theorem implies in our case that if a perturbation \eqref{eq:dispdef2} of {\it one} Hamiltonian of the Principal Hierarchy is integrable, then the involutivity condition \eqref{eq:dispinv} singles out an operator \eqref{doper}, which  order by order in $\e$ reconstructs the dispersive tail of {\it all} flows. This operator is uniquely defined, modulo total derivatives and the relations \eqref{flatfunct} defining the dispersionless Hamiltonian densities.

\subsection{The resolved conifold and the Ablowitz--Ladik hierarchy}
\label{sec:ablo}
Consider now the Principal Hierarchy for the resolved conifold in the equivariantly Calabi--Yau case $\lambda_1=-\lambda_2=\lambda$. In this case the prepotential is
\beq
F_0^{X_1, \rm ad}=-\frac{1}{2\lambda^2}v^2 w + \Li_3(e^w)
\label{eq:f0conad}
\eeq
and, from the fact that $\eta_{vv}=0$, the non-linear wave equation \eqref{eq:nlinconifold} has a vanishing rectangular term
\beq
w_{tt}=-\lambda^2 \l(\frac{e^w}{1-e^w} w_x\r)_x.
\eeq
This equation  was recognized in \cite{MR2462355} to be related to the dispersionless limit of the {\it Ablowitz--Ladik lattice} \cite{MR0377223}. We will here review, almost verbatim, the arguments of \cite{MR2462355} relating the solution of WDVV \eqref{eq:f0conad} to such an integrable lattice. The basic flow of the system is
\bea\label{abla1}
&&
i\,\dot a_n = -\frac12\,(1-a_n b_n) (a_{n-1}+a_{n+1})+ a_n,
\nn\\
&&
\\
&&
i\, \dot b_n = \quad\frac12\,(1-a_n b_n)(b_{n-1} + b_{n+1}) - b_n,
\nn
\eea
where $\{a_n, b_n :\bbZ \to \bbC\}$. Introducing new variables
\bea\label{abla6}
&&
u_n =-\log(1-a_n b_n),
\nn\\
&&
\\
&&
y_n = \frac1{2i} \left( \log{\frac{a_n}{a_{n-1}}} -\log{\frac{b_n}{b_{n-1}}} \right),
\nn
\eea
the evolution \eqref{abla1} can be written as a Hamiltonian flow generated by
\beq\label{abla7}
H_{\mathrm{AL}}=\sum_n \sqrt{\left(1-e^{-u_n}\right)\left(1-e^{-u_{n-1}}\right)}\, \cos y_n
\eeq
with the Poisson bracket
\beq\label{abla8}
\{ u_n, y_m\} = \delta_{n,m-1} -\delta_{n,m},\quad \{ u_n,u_m\}=\{ y_n, y_m\}=0.
\eeq
By taking the long-wave expansion we continuously interpolate the discrete dependent variables $u_n$, $t_n$ through functions $u(X,t)$, $y(X,t)$
\beq
u_n =u(\e n, \e  t), \quad y_n = y(\e  n, \e  t).
\eeq
This leads, at the leading order in $\e$, to the dispersionless system
\bea\label{abla9}
&&
u_t =\de_X\left[\left( e^{u}-1\right)\, \sin y\right],
\nn\\
&&
\\
&&
y_t\,= \de_X \left[ e^{-u} \cos y\right].
\nn
\eea
In order to make contact with the Principal Hierarchy of the resolved conifold, we will follow the argument of \cite{MR2462355} replacing $v(X)$, $y(X)$ by 
\bea\label{uw1}
x &:=& i \lambda X, \\
v(x)&:=& i y(x) \lambda, \\
w(x)&:=& \frac{i\e \lambda \de_x}{e^{i\e \lambda \de_x} -1}\, u(x).
\eea
In this way, the Poisson brackets of $w$ and $v$ take the standard form \eqref{eq:poissbrack}, and the Hamiltonian \eqref{abla7} becomes upon interpolation
\bea\label{eq:hal}
H_{\rm AL}&=& \int h_{\rm AL}\, \rd x \nn \\ &=& \int  \sqrt{ \left( 1-\exp\left\{ \frac{1-e^{i\e \lambda \de_x}}{i\e \lambda \de_x}\, w\right\}\right)
\left(1- \exp\left\{ \frac{e^{i\e \lambda \de_x}-1}{i\e \lambda \de_x}\, w\right\}\right)}\, \cosh\left( \frac{v}{\lambda}\right)\, \rd x, \nn \\
\eea
\beq
h_{\rm AL}=
\left(-1+e^{w}\right) \cosh \left(\frac{v}{\lambda }\right)-\frac{\left(e^{w} \lambda ^2 \cosh
   \left(\frac{v}{\lambda }\right) \left(4 \left(-1+e^{w}\right) w_{xx}-3 (w_{xx})^2\right)\right) \e ^2}{24
   \left(-1+e^{w}\right)}+O\left(\e ^4\right).
\label{eq:halpert}
\eeq
It turns out that the Ablowitz--Ladik lattice admits an infinite set of
conserved currents \cite{MR0377223}; as opposed to the Toda case, these
currents do not come straightforwardly from a bi-Hamiltonian recursion
associated to a local Poisson pencil, due to the non-existence of an Euler vector field for the prepotential \eqref{eq:f0conad}. It is easy to show that at the $\cO(\e^0)$ an infinite number of them\footnote{In fact all of them, with the sole exception \cite{MR2462355} of the one generating phase shifts of $a_n$, $b_n$.} coincide with the densities of the Principal Hierarchy associated to the prepotential \eqref{eq:f0conad}: the condition for a dispersionless density $f(v(x),w(x))$ to be in involution with the dispersionless Ablowitz--Ladik hamiltonian gives
\beq
\l\{H_{\rm AL}^{[0]} , \int_{S_1}f \r\} = 0 \Leftrightarrow \de^2_{ww}f+\l(\frac{\lambda^2 e^w}{1-e^w}\r)\de^2_{vv} f =0,
\label{eq:linweq}
\eeq
which is implied by \eqref{flatfunct} for $\a=\b$. \\

This connection provides us with a viable candidate hierarchy to relate to the Gromov--Witten theory of $X_1$ with $e_T(X_1)=0$, and led us to our Conjecture \ref{conj:AL} connecting the dispersionful Ablowitz--Ladik system with the all-genus theory of $X_1$. To this aim, and as a first step towards the reconstruction of the dispersionful hierarchy, let us remark here that by Theorem \ref{thm:dop}, Eq. \eqref{eq:hal} offers us a way to effectively construct the dispersive flows:

\begin{prop}
The $D$-operator for the Ablowitz--Ladik hierarchy reads at $\cO(\e^2)$
\bea
D_{\rm AL} f &=& f+ \e^2\Bigg[
\frac{e^{w(x)} \left(-1+2 e^{w(x)}\right) w'(x)^2
  f_{vv} \lambda ^4}{24
   \left(-1+e^{w(x)}\right)^2}+\frac{e^{w(x)} w'(x)^2
  f_{vvw} \lambda ^4}{12
   \left(-1+e^{w(x)}\right)} \nn \\ &+& \frac{e^{w(x)} w'(x) v'(x)
  f_{vvv} \lambda ^4}{6
   \left(-1+e^{w(x)}\right)}+\frac{v'(x)^2 f_{vv}
   \lambda ^2}{-12+12 e^{-w(x)}}+\frac{1}{12} v'(x)^2
  f_{vvw} \lambda ^2\Bigg] + \cO(\e^4). \nn \\
\label{eq:dop1loop}
\eea
\end{prop}
\noindent Eq. \eqref{eq:dop1loop} was obtained, with minor discrepancies due to a different choice of variables, in \cite{MR2462355}. A sketch of the proof can be found in Appendix \ref{sec:appdop}. Applying \eqref{eq:dop1loop} to the densities $h_{\a,p}$ of the Principal Hierarchy we find
\bea
h_v^{[2]}(v,w,v_x,w_x) &=& \frac{1}{24
   \left(-1+e^{w(x)}\right)^2} \Bigg[e^{w(x)} \lambda ^2 \left(-w(x)+2 e^{w(x)} (w(x)+1)-2\right) w'(x)^2 \nn \\
&-& 2 \left(-1+e^{w(x)}\right) \left(e^{w(x)} (w(x)-1)+1\right) v'(x)^2\Bigg] z^2+O\left(z^3\right), \\
h_w^{[2]}(v,w,v_x,w_x) &=& \frac{e^{w(x)} \left(\left(-1+2 e^{w(x)}\right) \lambda ^2 w'(x)^2-2 \left(-1+e^{w(x)}\right) v'(x)^2\right) z}{24 \left(-1+e^{w(x)}\right)^2} \nn \\ &+& \frac{e^{w(x)}}{24
   \left(-1+e^{w(x)}\right)^2} \Bigg[4
 \left(-1+e^{w(x)}\right) w'(x) v'(x) \lambda ^2  \nn \\ &+&  v(x)  \left(\left(-1+2 e^{w(x)}\right) \lambda ^2 w'(x)^2-2 \left(-1+e^{w(x)}\right) v'(x)^2\right)\Bigg] z^2+O\left(z^3\right). \nn \\
\label{eq:h1loop}
\eea
As an example, the leading order correction to \eqref{eq:nlinconifold} reads
\bea
w_{tt} &=& \l(\frac{-\lambda^2 e^{w}}{1-e^w}w_x\r)_x- \frac{e^{w(x)}}{24 \left(-1+e^{w(x)}\right)^4} \Bigg[\left(1+4 e^{w(x)}+e^{2 w(x)}\right) \lambda ^2 w'(x)^3 \nn \\ &+& \l(-2+2e^{2 w(x)}\r) \left(v'(x)^2-2 \lambda ^2 w''(x)\right) w'(x)-2   \left(-1+e^{w(x)}\right)^2 \nn \\ & \times & \left(2 v'(x) v''(x)-\lambda ^2 w^{(3)}(x)\right)\Bigg]_x\e^2+\cO(\e^4).
\eea
\subsection{Quasi-triviality and genus one Gromov--Witten invariants}
\label{sec:quasitr}
A key ingredient in the Dubrovin--Zhang analysis of bi-Hamiltonian evolutionary
hierarchies, which proved instrumental in their proof \cite{MR2092034} of the
$\bbP^1$/Toda correspondence, is the fact the hierarchy verifying Conjecture
\ref{conj:integrableGW} satisfies a {\it quasi-triviality} property:
\begin{defn}
A transformation of the form 
\beq
u^\a \to z^\a = u^\a+\sum_{k=1}^\infty \e^{2k} F_{k}^\a(\bfu, \bfu_x, \dots, \bfu^{(m(k))}),
\label{eq:quasimiura}
\eeq
with $m(k)$ a monotonically increasing, positive integer-valued sequence and $F_{k}^\a$ a degree $k$ rational function of $\bfu^{(j)}$ for $j>0$, will be called a quasi-Miura transformation. The hierarchy \eqref{eq:dispexp} is called quasi-trivial if there exists a quasi-Miura transformation reducing it, in the new variables, to its dispersionless $\e=0$ truncation.
\end{defn}
The quasi-Miura transformation \eqref{eq:quasimiura} is said to be $\tau$-symmetric if there exists a formal power series
\beq
\cF(\e, \bfu, \bfu_x, \dots) = \sum_{k=0}^\infty \e^{2k} \cF_{k}(\bfu, \bfu_x, \dots, \bfu^{(m(k)-2)})
\label{eq:Fvsde2F}
\eeq
where again  $\cF_{k}(\bfu, \bfu_x, \dots, \bfu^{(m(k)-2)})$ is a degree $k$ rational function of $\bfu^{(j)}$ for $j>0$, such that
\beq
F_k^\a(\bfu, \bfu_x, \dots, \bfu^{(m(k))})= \frac{\de^2 \cF_k}{\de x \de t_{\a,0}}(\bfu, \bfu_x, \dots, \bfu^{(m(k)-2)})
\label{eq:tausymmmiura}
\eeq
Comparing \eqref{eq:tausymmmiura} with \eqref{eq:tau} identifies $\cF_k$ as
the $\cO(\e^k)$ dispersive correction to the logarithm of the dispersionless $\tau$ function, i.e., in the case of the Principal Hierarchy \eqref{eq:princhier}, of the genus zero topological $\tau$-function of Theorem \ref{thm:dubr}. Conjecture \ref{conj:integrableGW} then states that precisely one such object  should correspond to the full descendent, all-genus Gromov--Witten potential. \\
\begin{rem}
It should be stressed that the fact that the Miura-type transformation \eqref{eq:quasimiura} is
quasi-Miura -- namely,  that $F_{k}^\a(\bfu, \bfu_x, \dots, \bfu^{(m(k))})$ is a
rational function of the jet variables -- is something not completely expected
within the realm of non-conformal Frobenius structures: in ordinary Gromov--Witten
theory the existence of an Euler vector field heavily enters the proof of
rationality of the Miura-type transformation \cite{dubrovin-2001}, and we have
no analogue for the Euler vector field here. For elliptic Gromov--Witten invariants,
however, the fact that the transformation is rational in the jet variables
goes through to the equivariant setting as well, as it is a consequence of the genus one
topological recursion relations; the only part where the homogeneity condition
intervenes is in fixing a one-parameter ambiguity in the $G$-function.  Our study
of the local $\bbC\bbP^1$/Ablowitz--Ladik correspondence suggests that this fact
should perhaps extend to more general examples also in higher genus, perhaps encompassing
all toric target spaces with equivariantly semi-simple quantum co-homology. 
\end{rem}
The rest of this section is devoted to the use of a quasi-triviality transformation to give a proof of Conjecture \ref{conj:AL} at $\cO(\e^2)$, that is, for $g = 1$ Gromov--Witten invariants. Our proof consists of the following three steps:
\bit
\item[(a)] proving that the Ablowitz--Ladik hierarchy is quasi-trivial at $\cO(\e^2)$;
\item[(b)] fixing a suitable choice of dependent variables leading to a $\tau$-symmetric transformation \eqref{eq:tausymmmiura} at $\cO(\e^2)$;
\item[(c)] proving that the logarithm of the $\tau$-function thus obtained coincides with the genus one, full descendent potential of $X_1$ with anti-diagonal action.
\eit
In the context of the usual theory of (conformal) semi-simple Frobenius
structures, step (a) is a consequence of the theory of $(0,n)$ Poisson pencils
on the loop space, while step (b) and (c) follow from the axiom of
linearization of Virasoro constraints. In absence of bi-Hamiltonianity and
possibly Virasoro-type constraints, we will need to perform steps (a)-(c) ``by
hand'', guided to some extent by the analogy with the Extended Toda hierarchy. \\

\noindent Let us start from step (a):

\begin{thm}
There exists an infinitesimal time-$\e$ canonical quasi-Miura transformation
\beq
u^\a \to z^\a = u^\a+\e \l\{u^\a, K\r\}+\cO(\e^2),
\label{eq:quasiK}
\eeq
where
\beq
K = -\frac{\e}{24}\int_{S^1} \Bigg[ F_+(x) \log
   F_+(x) + F_-(x) \log
   F_-(x) -  2 \log \left(-1+e^{w(x)}\right) v'(x)\Bigg] \rd x
\label{eq:K}
\eeq
and
\beq
F_\pm(x):= v'(x) \pm \sqrt{\frac{e^{w(x)}}{-1+e^{w(x)}}} \lambda 
   w'(x)
\eeq
which sends solutions of the Principal Hierarchy to those of its $\cO(\e^2)$ correction \eqref{eq:h1loop}.
\label{thm:K}
\end{thm}
\noindent To prove it we will make use of the following technical lemma from \cite{MR2462355}:
\begin{lem}
A density $h(\bfu, \bfu_x, \ldots)$, depending at most rationally on the jet variables $\bfu^{(n)}$ for $n>1$ is a total derivative $\de_x g(\bfu, \bfu_x, \ldots)$ if and only if
\beq
\frac{\d}{\d \bfu(y)} \int_{S^1} h(\bfu(x), \bfu_x(x), \ldots) \rd x=0.
\eeq
\label{lem:totalder}
\end{lem}
\noindent{\it Proof of Theorem \ref{thm:K}.} The proof follows from a very lengthy, but straightforward application of Lemma \ref{lem:totalder}. The Hamiltonian densities $h^{\rm qt}_{\a,p}(\bfz, \bfz_x, \bfz_{xx})$ obtained by composition of the dispersionless densities $h_{\a,p}(\bfu)$ with the $\cO(\e^2)$ quasi-Miura transformation \eqref{eq:quasiK} are given by
\beq
h^{\rm qt}_{\a,p}(\e, \bfz, \bfz_x, \bfz_{xx}, \dots) :=h_{\a,p}(\bfu(\e, \bfz, \bfz_x, \dots))=h_{\a,p}(\bfz)+\e\l\{h_{\a,p}, K\r\}+o(\e^2)
\eeq
since the trasformation generated by $K$ is canonical. In particular the $\cO(\e^2)$ correction is 
\beq
h^{[2], \rm qt}_{\a,p}(\bfz, \bfz_x, \bfz_{xx}) = \e\l\{h_{\a,p}, K\r\}.
\eeq
We claim that this reproduces the leading dispersive correction \eqref{eq:h1loop} of the dispersionless Ablowitz--Ladik flows. In the following we denote by $h^{\rm Dop}_{\a,p}$ the densities \eqref{eq:h1loop} we got by acting with the $D$-operator \eqref{eq:dop1loop}, to distinguish them from the ones obtained via \eqref{eq:quasiK}; accordingly, the corresponding generating functions will be written $h^{\rm Dop}_{\a}(z)$ and $h^{\rm qt}_{\a}(z)$. Define
\beq
r_{\a}(z) :=  h^{\rm qt}_{\a}(z)-h^{\rm Dop}_{\a}(z).
\eeq
We refrain here from reproducing the exact form of $r_{\a}(z)$, which would
occupy alone a few pages\footnote{We will be happy to provide the interested
  reader with the details of this calculation.}. A direct computation using \eqref{eq:dop1loop} and \eqref{eq:K} shows that
\bea
 \frac{\d}{\d \bfu(x)} \int_{S^1} r^{[2]}_{\a}(z)(\bfu, \bfu_y, \bfu_{yy}) \rd y = 
0.
\eea
Therefore we conclude that
\beq 
H^{\rm qt, [2]}_{\a,p}= H^{\rm Dop, [2]}_{\a,p}.
\label{eq:HqtHdop}
\eeq

\begin{flushright}$\square$\end{flushright}
\begin{rem}
Even though the Hamiltonian flows they generate are equal by \eqref{eq:HqtHdop}, the Hamiltonian densities $h^{\rm qt}_{\a,p}$ and $h^{\rm Dop}_{\a,p}$ differ greatly in form; in particular the quasi-Miura transformation \eqref{eq:quasiK} introduces a rational, rather than polynomial, dependence of $h^{\rm qt}_{\a,p}$ on the jet variables $\bfu^{(n)}$ for $n \geq 1$, which should disappear only at the level of the flows by \eqref{eq:HqtHdop}. For the basic dispersionless Hamiltonian of the Ablowitz--Ladik lattice we have for example
\beq
\l\{(1-e^w)\cosh\l(\frac{v}{\lambda}\r), K \r\} = -\frac{P(\bfu, \bfu_x, \bfu_{xx}, \bfu^{(3)})}{48
   \left(-1+e^{w(x)}\right)^2 \left(e^{w(x)} \lambda ^2 w'(x)^2-\left(-1+e^{w(x)}\right) v'(x)^2\right)^2}
\eeq
where $P(\bfu, \bfu_x, \bfu_{xx}, \bfu^{(3)})$ looks like

\bea 
P &=& e^{w(x)-\frac{v(x)}{\lambda }} \lambda ^2 \Bigg[\left(1+e^{\frac{2 v(x)}{\lambda }}\right) \Big(-3 e^{3 w(x)} \lambda ^4
   w'(x)^6+e^{w(x)} \left(-1+e^{w(x)}\right) \lambda ^2  \nn \\ &\times& \left(e^{w(x)} \left(1+2 e^{w(x)}\right) w''(x) \lambda ^2+\left(1+5
   e^{w(x)}\right) v'(x)^2\Big) w'(x)^4\r) -2 e^{w(x)} \left(-1+e^{w(x)}\right)^2  \nn \\
 &\times& \lambda ^2 \left(e^{w(x)} \lambda ^2    w^{(3)}(x)-2 v'(x)
   v''(x)\right) w'(x)^3-2 e^{w(x)} \left(-1+e^{w(x)}\right)^2  v'(x)^4+\dots
   \Bigg]. \nn \\
\eea
\end{rem}

It is particularly instructive to consider the result of the quasi-Miura transformation on the variable $w$. We find
\beq
\e \l\{w, K\r\}=-\lambda^2 \e^2 \frac{\de^2\tilde\cF_1(w, v_x, w_x)}{\partial x^2},
\label{eq:quasitrw}
\eeq
where
\beq
\tilde\cF_1(w, v_x, w_x):=\frac{1}{24}\log\l( v'(x)^2+\frac{\lambda^2 e^{w(x)}}{1-e^{w(x)}}w'(x)^2\r)+\frac{1}{12}\Li_1(e^{w(x)}).
\label{eq:F1tilde}
\eeq
On the other hand, for the $v$ variable we find
\beq
\e \l\{v, K\r\}=-\lambda^2 \e^2 \l(\frac{\de^2\tilde\cF_1(w, v_x, w_x)}{\partial x \partial t^{2,0}}+ \frac{1}{6} \frac{\de^2 \Li_1(e^{w})}{\partial x \partial t^{2,0}} \r) \neq -\lambda^2 \e^2 \frac{\de^2\tilde\cF_1(w, v_x, w_x)}{\partial x \partial t^{2,0}},
\label{eq:quasitrv}
\eeq
that is, the quasi-triviality transformation \eqref{eq:quasiK} is {\it not}
$\tau$-symmetric. Indeed, except for the KdV case, the incompatibility between
canonicity and $\tau$-symmetry of the quasi-Miura transformation seems to be a
fairly generic fact\footnote{This point was strongly emphasized to us in an
  enlightening discussion with B.~Dubrovin.}, also for bi-Hamiltonian systems
like the Extended Toda hierarchy. Quite interestingly, the similiarity with
the case of the Gromov--Witten theory of $\bbP^1$ and the Extended Toda
hierarchy is even more striking, as the discrepancy between the transformation
generated by $K$ and the one via the logarithm of a dispersive $\tau$-function
is present in only {\it half} of the change of variables, and it is equal to
twice the term independent of space derivatives (in the language of
\cite{Dubrovin:1997bv}, the $G$-function) inside $\tilde\cF_1$. \\

It should be
emphasized that the form of the $G$-function in the Toda case relies on the
existence of a grading condition for the Frobenius manifold associated to
$QH^\bullet(\bbP^1)$, which in particular allows to fix the degree zero terms at
$g=1$ \cite{Dubrovin:1997bv}. As we will see, in our case this is achieved by shifting the
form of the $g=1$ full-descendent free energy by the constant map term 
\beq
\langle \tau_0(1) \rangle^X_{1,1,0}=- \int_{\cM_{1,1}} \lambda_1=-\frac{1}{24},
\eeq
where $\lambda_1=c_1(\mathbb{E})$ is the Chern class of the Hodge
line bundle on $\overline{\cM}_{1,1}$, through
\beq
\tilde \cF_1 \to \cF_1:=\tilde \cF_1 -  \frac{w(x)}{24}.
\label{eq:dzeroshift}
\eeq
Inspired by the analogy with the $\bbP^1$/Toda case, we are then led to consider the following $\tau$-symmetric ansatz for the choice of dependent
variables at $\cO(\e^2)$:
\beq
u_\a \to u_\a + \e^2 \frac{\de^2\cF_1(w, v_x, w_x)}{\partial x \partial t^{\a,0}}+\cO(\e^4),
\label{eq:tausymf1}
\eeq
where
\beq
\cF_1=\frac{1}{24}\log\l( v'(x)^2+\frac{\lambda^2
  e^{w(x)}}{1-e^{w(x)}}w'(x)^2\r)+\frac{1}{12}\Li_1(e^{w(x)})-  \frac{w(x)}{24}.
\label{eq:F1}
\eeq

We are now in position to prove Theorem \ref{thm:main}: by the definition of quasi-triviality, the $\cO(\e^2)$ correction to the logarithm of the $\tau$-function $\cF_0$ in \eqref{eq:dercorr} should be obtained by plugging into \eqref{eq:F1} the solution of the Principal Hierarchy with initial data \eqref{eq:initial}; the result of the composition will be again denoted by the same symbol $\mathcal{F}_1(\mathbf{t})$. Accordingly, we introduce genus 1 ``big correlators''
\beq
\bra\bra \tau_{p_1}(\phi_{\alpha_1}), \dots, \tau_{p_k}(\phi_{\a_k}) \ket\ket_1 := \frac{\de^{k} \mathcal{F}_1(\mathbf{t})}{\de{t^{\alpha_1,p_1}} \dots  \partial {t^{\a_k,p_k}}}.
\label{eq:dercorr1loop}
\eeq
\noindent{\it Proof of Theorem \ref{thm:main}.} The statement for $g=0$ was
proven in Sec. \ref{sec:dispconif} (Theorem \ref{thm:dubr} and Proposition
\ref{prop:norm}). For $g=1$, notice that the descendent Gromov--Witten
invariants of $X_1$ are completely determined in a recursive fashion by the
following two formulas: the first is the $g=1$ case of the higher genus
multi-covering formula found in \cite{Marino:1998pg, Gopakumar:1998jq,
  Katz:1999xq, MR1728879, Bryan:2004iq} that yields the primary potential of $X_1$ as
\beq
F_1^{X_1}(t^{2,0})=- \frac{t^{2,0}}{24} + \sum_{d=1}^\infty N^{(1)}_{1,d} e^{d t^{2,0}}=- \frac{t^{2,0}}{24}+\frac{1}{12}\Li_1(e^{t_{2,0}}).
\label{eq:primpotg1}
\eeq
The second is the set of $g=1$ topological recursion relations \cite{Dijkgraaf:1990nc, Kontsevich:1997ex, MR1672112}
\bea
\bra\tau_{p}(\phi_\a) \ket_{1,1,d}^{X_1} &=&
\sum_{d_1+d_2=d}\bra\tau_{p-1}(\phi_\a) \tau_{0}(\phi_\nu)
\ket_{0,2,d_1}^{X_1} \eta^{\mu \nu}\bra\tau_{0}(\phi_\mu) \ket_{1,1,d_2}^{X_1}
\nn \\ &+& \frac{1}{24} \eta^{\mu \nu}\bra\tau_{p-1}(\phi_\a) \tau_{0}(\phi_\nu) \tau_{0}(\phi_\mu)  \ket_{0,3,d}^{X_1},
\label{eq:g1trr}
\eea
that fully determine genus one descendent invariants in terms of the genus one primaries and genus zero descendents.
We will prove here that the Ablowitz--Ladik $\tau$-function \eqref{eq:F1} implies both \eqref{eq:primpotg1} and \eqref{eq:g1trr}. \\

\noindent Consider first the small phase reduction of $\cF(\mathbf{t})$, i.e.
\bea
u^\a(\mathbf{t})\l|_{\substack{t^{\a,p}=0 \\ \hbox{\footnotesize for } p>0}}\r. &=& t^{\a,0}+\delta^{\a,0}x. 
\label{eq:initcond}
\eea
Replacing into \eqref{eq:F1} we find
\bea
\cF_1(\mathbf{t})\l|_{\substack{t^{\a,p}=0 \\ \hbox{\footnotesize for }
    p>0}}\r. &=&  - \frac{t^{2,0}}{24}+\frac{1}{12}\Li_1(e^{t^{2,0}}), 
\eea
which proves \eqref{eq:primpotg1}. \\

\noindent To see that $\cF_1(\mathbf{t})$ correctly embeds the full-descendent information too, we follow \cite{Dubrovin:1997bv} and compute
\bea
\cC_{\a,p} &:=& \frac{\de \cF_1}{\de t^{\a,p}}-\frac{\de^2 \cF_0}{\de t^{\a,p-1} \de t^{\nu,0}}\eta^{\mu\nu}\frac{\de \cF_1}{\de t^{\mu,0}} \nn \\ &=&  
\sum_{\gamma=1,2}\l[\frac{\de \cF_1}{\de u^\gamma} \frac{\de u^\gamma}{\de t^{\a,p}}+\frac{\de \cF_1}{\de u_x^\gamma} \de_x \l(\frac{\de u^\gamma}{\de t^{\a,p}}\r)  - \de_\nu h^{[0]}_{\a,p-1}\eta^{\mu\nu}\l(\frac{\de \cF_1}{\de u^{\g}}\frac{\de u^\g}{\de t^{\mu,0}}+ \frac{\de \cF_1}{\de u_x^{\g}}\de_x\frac{\de u^\g}{\de t^{\mu,0}} \r)\r] \nn \\
&=& \sum_{\gamma=1,2}\Bigg[\frac{\de \cF_1}{\de u^\gamma} \de_\nu  h^{[0]}_{\a,p-1} \frac{\de u^\gamma}{\de t^{\mu,0}}\eta^{\mu \nu}+
 \frac{\de \cF_1}{\de u_x^\gamma} \de_x \l( \de_\nu h^{[0]}_{\a,p-1} \frac{\de u^\gamma}{\de t^{\mu,0}}\eta^{\mu \nu}\r) \nn \\
&-& \de_\nu h^{[0]}_{\a,p-1}\eta^{\mu\nu}\l(\frac{\de \cF_1}{\de u^{\g}}\frac{\de u^\g}{\de t^{\mu,0}}+ \frac{\de \cF_1}{\de u_x^{\g}}\de_x\frac{\de u^\g}{\de t^{\mu,0}} \r)\Bigg] \nn \\
&=&  \sum_{\gamma=1,2}\frac{\de \cF_1}{\de u^\gamma_x} \de_x \de_\nu h^{[0]}_{\a,p-1} \frac{\de u^\gamma}{\de t^{\mu,0}}\eta^{\mu \nu},
\label{eq:Cap}
\eea
where in the second line we used the chain rule and the fact that
\beq
\de_\nu h^{[0]}_{\a,p-1} = \bra\bra \tau_{p-1} (\phi_\a) \tau_{0} (\phi_\nu) \ket\ket_0 = \frac{\de^2 \cF_0}{\de t^{\a,p-1} \de t^{\nu,0}},
\eeq
while in the third line we used the genus zero topological recursion relations \eqref{eq:dubrTRR}
\bea
\frac{\de u^\a}{\de t^{\b,q}} &=& \eta^{\a\g}\frac{\de^3 \cF_0}{\de t^{\g,0} \de t^{1,0} \de t^{\b,q}} =  \eta^{\a\g}\frac{\de^2 \cF_0}{ \de t^{\d,0} \de t^{\b,q-1}} \eta^{\d \e}\frac{\de^3 \cF_0}{\de t^{\g,0} \de t^{1,0} \de t^{\d,0}} \nn \\
&=& \frac{\de u^\a}{\de t^{\d,0}}\eta^{\d \e} \de_\e h^{[0]}_{\a,p-1}.
\eea
Moreover Eq. \eqref{eq:firstflowsconifold} with $\lambda_1=-\lambda_2=\lambda$  and the explicit form \eqref{eq:F1} yield
\bea
 \sum_{\gamma=1,2} \frac{\de \cF_1}{\de u^\gamma_x}  \frac{\de u^\gamma}{\de t^{\mu,0}}\eta^{\mu \nu} &=& \frac{\de \cF_1}{\de v_x}\frac{\de v}{\de t^{\nu,0}}+\frac{\de \cF_1}{\de w_x}\frac{\de w}{\de t^{\nu,0}} =\frac{\eta^{\mu \nu}}{24} \frac{2 v'(x) \de_{t^{\mu,0}} v + 2 \frac{\lambda^2 e^{w(x)}}{1-e^{w(x)}} w'(x) \de_{t^{\mu,0}} w}{v'(x)^2 + \frac{\lambda^2 e^{w(x)}}{1-e^{w(x)}}w'(x)^2} \nn \\
 &=&-\frac{\lambda^2}{12} \delta^{\nu,0}.
\eea
This implies
\beq
\cC_{\a,p}=-\frac{\lambda^2}{12} \partial_x \partial_w h^{[0]}_{\a,p} =-\frac{\lambda^2}{12}\frac{\de^3 \cF_0}{\de t^{1,0} \de t^{2,0} \de t^{\a,p-1}}=\frac{1}{24}\eta^{\mu\nu} \frac{\de^3 \cF_0}{\de t^{\a,p-1} \de t^{\nu,0} \de t^{\mu,0}}.
\eeq
Combining the last equality with the definition of $\cC_{\a,p}$ in the first line of \eqref{eq:Cap} we obtain
\beq
\bra\bra\tau_{p}(\phi_\a) \ket\ket_1 = \bra\bra\tau_{p-1}(\phi_\a) \tau_{0}(\phi_\nu)  \ket\ket_0 \eta^{\mu \nu}\bra\bra\tau_{0}(\phi_\mu) \ket\ket_1+\frac{1}{24} \eta^{\mu \nu}\bra\bra\tau_{p-1}(\phi_\a) \tau_{0}(\phi_\nu) \tau_{0}(\phi_\mu)  \ket\ket_0,
\eeq
which, setting $t^{\a,p}=0$ for $p>0$ in \eqref{eq:dercorr1loop} and expanding in $e^{t_{2,0}}$, implies \eqref{eq:g1trr}.
\begin{flushright}$\square$\end{flushright}

\subsection{A look at the higher genus theory}
\label{sec:g2}
A natural further step would be to generalize the results of the previous
section to higher genus Gromov--Witten invariants. As usual, however, the
degree of difficulty undergoes a phase transition as soon as  $g>1$, and the search of straightforward generalizations of the methods we used becomes unwieldy. In particular, the construction of the quasi-triviality transformation appears to be very hard, let alone finding a suitable $\tau$-symmetric form to compare with the higher genus Gromov--Witten potentials. \\

However, there is still room for a number of non-trivial tests of Conjecture \ref{conj:AL}. To see this, recall that in the previous section we found three ways to construct the $\cO(\e^2)$ dispersive tail of the Ablowitz--Ladik hierarchy:
\ben[(i)]
\item via the $D$-operator \eqref{eq:dop1loop};
\item via the canonical quasi-Miura transformation \eqref{eq:quasiK};
\item via the $\tau$-symmetric quasi-Miura transformation \eqref{eq:tausymf1}.
\een
(i) and (ii) are equivalent by Theorem \ref{thm:K}, and (ii) and (iii) because
rational Miura transformations form a group. At the level of the flows the
statement of Theorem \ref{thm:K} is stronger than that, meaning that this
equivalence is realized as an equality of the form of the equations of the
hierarchy \eqref{eq:HqtHdop}. By \eqref{eq:quasitrw} and \eqref{eq:quasitrv}
this is not the case for (ii) and (iii), where 
$\tau$-symmetry is broken in the canonical setup; moreover, the canonical free
energy $\tilde\cF$ such that $w^{[\rm d-op] }(\mathbf{t})=\de^2_x
\tilde{\cF}(\mathbf{t})$ coincides with the topological free energy $\cF$ up
to a Miura transformation, consisting in a shift by terms
whose restriction to primary invariants involves only degree zero terms.
The situation is schematized in Table \ref{tab:disp}. \\

\begin{table}[h]
\begin{tabular}{ccccc}
\bf D-operator & & \bf Canonical q.t. & & \bf $\tau$-symmetric q.t. \\
\hline
\\
$w^{[\rm d-op]}(\mathbf{t}, \e)$ & $=$ & $w^{[\rm c.q.t.]}(\mathbf{t}, \e)$ & $=$ & $w^{[\rm \tau-s.q.t.]}(\mathbf{t}, \e)|_{d>0}$ \\
$v^{[\rm d-op]}(\mathbf{t}, \e)$ & $=$ & $v^{[\rm c.q.t.]}(\mathbf{t}, \e)$ & $\neq$ & $v^{[\rm \tau-s.q.t.]}(\mathbf{t}, \e)_{d>0} $
\end{tabular}
\vspace{.5cm}
\caption{\small Relations between solutions of the dispersionful Ablowitz--Ladik
  hierarchy at $\cO(\e^2)$. The equality between the second and the third
  column holds up to a Miura transformation whose restriction to the
  small phase space involves only degree zero terms.}
\label{tab:disp}
\end{table}

The objects to construct for the purpose of computing higher genus Gromov--Witten invariants are  $v^{[\rm \tau-s.q.t.]}(\mathbf{t}, \e)$ and $w^{[\rm \tau-s.q.t.]}(\mathbf{t}, \e)$ at higher order in $\cO(\e)$; as we emphasized, the relevant $\tau$-symmetric quasi-Miura transformation seems however very difficult to obtain. On the other hand, focusing on the first line we see that at the leading order in the perturbative expansion in $\e$ we have
\beq
w^{[\rm d-op]}(\mathbf{t}, \e) = w^{[\rm \tau-s.q.t.]}(\mathbf{t}, \e) = \frac{\de^2 \cF}{\de x^2}(\mathbf{t})+ \cO(\e^4)
\label{eq:wdopwtsqt0}
\eeq
up to terms related to constant maps contribution upon restriction to primary
fields, as in \eqref{eq:dzeroshift}. In the following we will make the following important
\begin{assm}
The equality \eqref{eq:wdopwtsqt0} holds true at genus $\cO(\e^4)$:
\beq
w^{[\rm d-op]}(\mathbf{t}, \e) = w^{[\rm \tau-s.q.t.]}(\mathbf{t}, \e) = \frac{\de^2 \cF}{\de x^2}(\mathbf{t})+ \cO(\e^6)
\label{eq:wdopwtsqt}
\eeq
up to a quasi-Miura transformation whose restriction to primaries is
determined by {\rm degree zero} Gromov--Witten invariants.
\label{assm:tausymm}
\end{assm}
That is, even if we do not know the form of $w^{[\rm d-op]}(\mathbf{t}, \e)$
as a rational function in the derivatives of the fields beyond $\cO(\epsilon^2)$, we assume that it is a double
derivative of some (rational) local functional $\cF$. In our situation, we
have little guidance for the construction of the right quasi-Miura
transformation which determines the form of $\cF$ at higher genus, but on the other hand
computing higher order corrections to the $D$-operator \eqref{eq:dop1loop},
and therefore to $w^{[\rm d-op]}(\mathbf{t})$, is just a matter of
computational time and stamina. Indeed the involutivity condition
\eqref{eq:dispinv}  gives a self-contained way to find $w^{[\rm
    d-op]}(\mathbf{t}, \e)$ at any order in $\e$, thus  allowing a complete
recursive reconstruction of the $\e$ expansion of the flows. As for the genus
one case, we might be missing a possible contribution from constant maps here; the computation below
will indeed be insensitive to degree zero invariants.\\

\noindent {\it Proof of Theorem \ref{cor:g=2}:} we will work out the
consequences of Theorem \ref{thm:dop} at $\cO(\e^4)$ by putting
$H_{\a,p}=H_{\rm AL}$, where the $\cO(\e^4)$ expansion of $H_{\rm AL}$ was given in \eqref{eq:halpert}. This determines the two-loops $D$-operator of the Ablowitz--Ladik lattice; for future use we give here the expression of the coefficients $b^{[4]}_{vv}$ and $b^{[4]}_{vvv}$
\beq
D^{[4]}f = b^{[4]}_{vv} f_{vv} + b^{[4]}_{vvv} f_{vvv} + \dots
\eeq
where, up to a total derivative, we have
\bea
\label{eq:MISTIMO0}
b^{[4]}_{vv} &=& \frac{\lambda ^2}{{5760 \left(-1+e^{w(x)}\right)^4}} \Bigg[
-8 e^{w(x)} \left(-1+e^{2 w(x)}\right) v'(x)^4+4 \left(-1+e^{w(x)}\right)  \Big(e^{w(x)} \nn \\ & \times &\left(11+7 e^{w(x)}\right)-2\Big)
   \lambda ^2 w''(x) v'(x)^2-8 \left(-1+e^{w(x)}\right) \left(e^{w(x)} \left(-1+19 e^{w(x)}\right)+2\right)  \lambda ^2 \nn \\ & \times & w'(x)
   v''(x) v'(x)+\lambda ^2  \Big(\left(4-e^{w(x)} \left(15 e^{w(x)} \left(-4+e^{w(x)}\right)+19\right)\right) \lambda ^2 w''(x)
   w'(x)^2 \nn \\ &+&  8 e^{w(x)} \left(-1+e^{w(x)}\right) \Big(\left(e^{w(x)} \left(-7+3 e^{w(x)}\right)+1\right) \lambda ^2
   w''(x)^2+\left(-1+e^{w(x)}\right) \nn \\ & \times & \left(5+7 e^{w(x)}\right) v''(x)^2\Big)\Big),
\Bigg] \\
b^{[4]}_{vvv} &=& \frac{1}{17280} \Bigg[\frac{\lambda ^4}{\left(1-e^{w(x)}\right)^3}\Bigg(
2 \Big(4 w(x) \left(-1+e^{w(x)}\right)^3+3 \Big(4 \log \left(-1+e^{w(x)}\right)
\nn \\ & \times &
   \left(-1+e^{w(x)}\right)^3+e^{w(x)} \left(e^{w(x)} \left(-107+32 e^{w(x)}\right)+131\right)-46\Big)\Big) w'(x) v'(x)
   w''(x) \lambda ^2\Bigg) \nn \\ &+& \left(\frac{3 \left(e^{w(x)} \left(-119+183
   e^{w(x)}\right)+46\right)}{\left(-1+e^{w(x)}\right)^3}-12 \log \left(-1+e^{w(x)}\right)-4 w(x)\right) w'(x)^2 v''(x) \lambda
   ^2 \nn \\ &-& \frac{144 e^{w(x)} \left(1+e^{w(x)}\right) w''(x) v''(x) \lambda ^2}{\left(-1+e^{w(x)}\right)^2}-12 \Bigg(-\frac{3
   \left(-5+9 e^{w(x)}\right)}{\left(-1+e^{w(x)}\right)^2}-2 \log \left(-1+e^{w(x)}\right) \nn \\ &+& 6 w(x)\Bigg) v'(x)^2   v''(x)\Bigg].
\label{eq:MISTIMO}
\eea

As an application of \eqref{eq:dop1loop} and \eqref{eq:MISTIMO}, consider the $t^{2,1}$ flow generated by $H_{2,1}$. For reasons that will be clear in a moment, we would like to compute the solutions $v^{[\rm d-op]}(t^{1,0}, t^{2,0}, t^{2,1}, \e)$ and $w^{[\rm d-op]}(t^{1,0}, t^{2,0}, t^{2,1}, \e)$, with all other times set to zero, of the $t^{2,1}$ flow. This will lead us to a proof of Theorem \ref{cor:g=2}. \\

\noindent From \eqref{eq:genintA}, \eqref{eq:unnormff} and \eqref{eq:cv1}-\eqref{eq:cw2} we have that
\beq
h^{[0]}_{2,1}(v,w)=-\frac{v^3}{6 \lambda^2}+v \Li_2(e^w)
\eeq
and therefore
\beq
h^{\rm d-op}_{2,1} := D_{AL} h^{[0]}_{2,1}=h^{[0]}_{2,1}+ \e^2 h^{[2]}_{2,1} + \e^4 h^{[4]}_{2,1}+\cO(\e^6),
\eeq
where 
\bea
h^{[2]}_{2,1} &=& \frac{e^{w(x)}}{24 \left(-1+e^{w(x)}\right)^2} \Bigg[4 \left(-1+e^{w(x)}\right) w'(x) v'(x) \lambda ^2+v(x) \Big(\left(-1+2 e^{w(x)}\right) \lambda ^2 w'(x)^2 \nn \\ &-& 2 \left(-1+e^{w(x)}\Big)    v'(x)^2\right)\Bigg], \\
h^{[4]}_{2,1} &=& -\frac{1}{\lambda^2} \l(v b^{[4]}_{vv} + b^{[4]}_{vvv} \r),
\eea
while on the other hand
\beq
h^{\rm d-op}_{1,0}=h^{[0]}_{1,0} = -\frac{v w}{\lambda^2}.
\eeq
Let us solve the dispersive equations 
\beq 
\frac{\de u^\a}{\de t^{2,1}} = \l\{u^{\a},\int_{S^1} h^{\rm
  d-op}_{2,1}(v,w)\r\}
\eeq
perturbatively in $t^{2,1}$ with the topological Cauchy datum \eqref{eq:initcond}. We find
\bea
w^{[\rm d-op]}(x, t^{2,0},t^{2,1})&=& t^{2,0}+t^{2,1} x+(t^{2,1})^2 \lambda ^2 \log
   \left(1-e^{t^{2,0}}\right) + \frac{e^{t^{2,0}} (t^{2,1})^3 x \lambda ^2}{-1+e^{t^{2,0}}}+\dots \nn \\ &+& \left(-\frac{e^{t^{2,0}} (t^{2,1})^2 \lambda ^2}{12 \left(-1+e^{t^{2,0}}\right)^2}+\frac{e^{t^{2,0}} \left(1+e^{t^{2,0}}  \right) (t^{2,1})^3 x \lambda ^2}{12
   \left(-1+e^{t^{2,0}}\right)^3} + \dots\right) \epsilon ^2 \nn \\ &+&  \left(-\frac{e^{t^{2,0}} \left(1+4 e^{t^{2,0}}+e^{2 t^{2,0}}\right) (t^{2,1})^2 \lambda ^2}{240 \left(-1+e^{t^{2,0}}\right)^4} + \dots\right) \epsilon^4+ \cO(\e^6).
\label{eq:wsol2l}
\eea
From now on we put $t^{2,0}=:t$. The last line of \eqref{eq:wsol2l} and the assumption \eqref{eq:wdopwtsqt} combined together lead to 
\beq
\frac{\de^4 \cF_2}{\de x^2 \de (t^{2,1})^2}\Bigg|_{\substack{t^{\a,p}=0 \\ \hbox{\footnotesize for } p>0}}=\frac{e^{t} \left(1+4 e^{t}+e^{2 t}\right) }{120 \left(-1+e^{t}\right)^4}=\frac{1}{120}\Li_{-3}\l(e^{t}\r).
\eeq
In Gromov--Witten theory the left hand side would represent the small phase correlator $\bra\bra \phi_1, \phi_1, \tau_1 (\phi_2), \tau_1 (\phi_2) \ket\ket^{X_1}_{2, \mathrm{sps}}$, where we define
\beq
\bra \bra \tau_{p_1}(\phi_{\a_1}) \dots \tau_{p_k} (\phi_{\a_k}) \ket \ket^{X_1}_{g, \mathrm{sps}}(x,t) := 
\sum_{d,n \geq 0} \frac{1}{n!} \bra  \tau_{p_1} (\phi_{\a_1}), \dots \tau_{p_k} (\phi_{\a_k}),  \overbrace{ x\phi_1 + t \phi_2 \dots  x\phi_1 + t\phi_2}^{\text{$n$  times}} \ket_{g,n+k,d}^{X_1}.
\eeq
Applying twice the puncture equation to $\bra\bra \phi_1, \phi_1, \tau_1 (\phi_2), \tau_1 (\phi_2) \ket\ket^{X_1}_{2, \mathrm{sps}}$ we can kill the two descendent insertions and reduce to the double derivative of the primary potential
\beq
2 \sum_{d \geq 0} \sum_{n=0}^\infty \frac{(t)^n}{n!} \bra   \phi_2,  \phi_2, \overbrace{ \phi_2, \dots,  \phi_2}^{\text{$n$  times}} \ket_{2,n+2,d} = 2 \frac{\de^2 F_2(t)}{\de t^2},
\eeq
that is
\beq
\frac{\de^2 F_2(t)}{\de t^2} = \frac{1}{240}\Li_{-3}\l(e^{t}\r).
\label{eq:der2g2}
\eeq
This reproduces exactly the higher genus formula for primary Gromov--Witten
invariants of $X_1$ \cite{Marino:1998pg, Gopakumar:1998jq, Katz:1999xq,
  MR1728879, Bryan:2004iq}
\beq
F_g^{X_1}(t)=\sum_{d=0}^\infty N^{(1)}_{g,d} e^{d t}=\frac{\l|B_{2g}\r|}{2g(2g-2)!}\Li_{3-2g}(e^{t})+\frac{\l|B_{2g} B_{2g-2}\r|}{2g (2g-2)(2g-2)!}
\label{eq:primpotggen}
\eeq
at genus 2, up to the constant map contribution, and proves Theorem \ref{cor:g=2}. \begin{flushright}$\square$\end{flushright}

\subsection{Higher descendent flows and the Ablowitz--Ladik equations}
By the same token, the complete solution $w(\mathbf{t})=\de^2_x \tilde{\cF}(\mathbf{t})$ of all flows should
contain information on descendent invariants; however, the discrepancy between
$\tilde{\cF}(\mathbf{t})$ and $\cF(\mathbf{t})$, which amounts to constant map terms when restricted
to primaries, might also affect positive degree invariants when it comes to
computing descendents. In particular the terms of  $\cO(t^{2,1})^{n+2}$ of
\eqref{eq:wsol2l} compute the right genus 2 Gromov--Witten invariants with single descendent
insertions at $n$ points only if $n\leq 2$. As for the genus one case, the precise
choice of dependent variables for the Ablowitz--Ladik equations is then crucial for
the computation of Gromov--Witten invariants, and in particular it should
correct the hydrodynamic Poisson structure \eqref{eq:poissbrack}, which is
left invariant by construction in the $D$-operator formalism, by higher order terms in $\epsilon$.\\

It is nonetheless remarkable that the dispersive Ablowitz--Ladik flows in the
$D$-operator form satisfy a number of constraints induced from the topology of
moduli spaces of stable maps. 
As an example, a little experimentation at the next few orders in $t^{2,1}$
shows that
\beq
\label{eq:exdesc1}
\frac{1}{n!} \frac{\de w^{[\rm d-op]}(x, t^{2,0},t^{2,1})}{\de (t^{2,1})^n}\bigg|_{t^{2,1}=0}= 
\sum_{k=0}^{n} a''_{k,n}(t) x^k
\eeq
with
\beq
a_{k,n}(t)=\l(\bary{c}n \\ k\eary\r) \frac{\de^{k} a_{0,n-k}(t)}{\de t^k}
\label{eq:stringrel}
\eeq
It is noteworthy that the relation \eqref{eq:stringrel}, which in
Gromov--Witten theory would be a consequence of the string axiom, is realized
by the dispersive Ablowitz--Ladik flows; we checked this up to
$\cO((t^{2,1})^7)$ (i.e. $n\leq 5$). Along the same lines, it is straightforward to switch on the $t^{1,1}$-flow of the Ablowitz--Ladik hierarchy and see that the dilaton constraint is satisfied too. As an example, for the $\cO((t^{2,1})^2)$, $\cO(\epsilon^{2g})$ coefficient $\tilde w^{(2)}_g(t^{1,1},t)$ of $w(x,t,t^{1,1},t^{2,1})$ we can give a closed expression for its $t^{1,1}$ dependence
\bea
\label{eq:exdesc3}
\tilde w^{(2)}_g &:=& \sum_{ n\geq 0} \frac{(t^{1,1})^n}{n!} \bra\bra \phi_1,\phi_1, \tau_{1}\phi_2, \tau_{1}\phi_2 , \overbrace{\tau_1\phi_1, \dots, \tau_1\phi_1}^{n \hbox{ \footnotesize \rm times}} \ket\ket^{X_1}_{g, \mathrm{sps}} \nn \\
&=& 
\left(\frac{1}{1-t^{1,1}}\right)^{2g+2}
\frac{\de^2}{\de y^2} F_g(y)\Bigg|_{y=\left(\frac{t}{1-t^{1,1}}\right)}, \qquad g=0,1,2
\eea
and it is immediate to see that the small-phase space dilaton equation holds
\beq
\left[(1-t^{1,1})\frac{\de}{\de t^{1,1}}-t \frac{\de}{\de t}-2-2g\right] \tilde w^{(2)}_g = 0.
\eeq

\bibliography{miabiblio}
\bibliographystyle{amsalpha}

\begin{appendix}

\section{Dispersive expansion of the Ablowitz--Ladik hierarchy}
\label{sec:appdop}
We collect here the details of the reconstruction of the dispersive tail of the dispersionless Ablowitz--Ladik hierarchy.
\subsection{Normal form for the $D$-operator}
Since the $D$-operator \eqref{doper} maps densities to densities,  the Hamiltonian flows it induces would be unmodified by the addition of a total derivative

\beq
Df\to \tilde{D} f = D f +g'
\label{eq:totder}
\eeq

Moreover, since such densities are supposed to integrate to Hamiltonians of a dispersionless hierarchy, they will be bound to satisfy a linear wave equation of the form \eqref{eq:linweq}. \\

Let us then give a normal form for the $D$-operator which solves this constraints. First of all, it was shown in \cite{MR2462355} that for systems of the type \eqref{eq:linweq}, the coefficients $ b^{[k]}_{l,m}$ in \eqref{doper} can be taken to be independent of $v$
\beq
 b^{[k]}_{l,m} (v_x, \dots, v^{(k)}, w \dots, w^{(k)})
\label{eq:bk1}
\eeq
up to a total derivative.  Let $I \in \bbN^{2k}$ be such that
\beq
\sum_{j=1}^{2k}\l[\frac{j+1}{2}\r] I_j=k
\eeq
 The differential polynomial $ b^{[k]}_{l,m}$ explicitly reads
\beq
 b^{[k]}_{l,m} =\sum_{I} d_{I,l,m}(w) \prod_{j=1}^{k}(v^{(j)}(x))^{I_{2j-1}} (w^{(j)}(x))^{I_{2j}}
\eeq
It is easy to realize that terms with $I_j=0$ for $j > [(k+1)/2]$ can be set to zero upon adding a suitable total derivative. The same is true for all remaining terms with $I_j=1$ and $I_{j-2}=(k-[(j+1)/2])/[(j-1)/2]$ for $2<j\leq [(k+1)/2]$.
This fixes entirely the ambiguity \eqref{eq:totder}. Furthermore, we can take into account \eqref{eq:linweq} by constraining $m \geq1 $; moreover, $\e$-exactness of the Hamiltonian $H_{1,0}$ generating the space translations sets $n>1$. We will take this as our normal form for the $D$-operator. The number of independent coefficients $N_k$ thus obtained at fixed $l$ and $m$, for the first few values of $k$, is $N_2=3$, $N_3=6$, $N_4=10$. 

\subsection{Computing the $D$-operator}
Let us then give an example of how to compute the $D$-operator by outlining the computation of the 1-loop case for the Ablowitz--Ladik hierarchy. Let $f$ be an arbitrary dispersionless Hamiltonian density \eqref{eq:linweq}. Then the $D^{[2]}$ correction to the $D$-operator should come from the $\cO(\e^2)$ involutivity condition
$$\l\{H^{[0]}_{AL}+\e^2 H^{[2]}_{AL}, \int_{S^1} (f+\e^2 D^{[2]}  f) \rd x \r\}=o(\e^2)$$
i.e., at the level of the densities and using Lemma \ref{lem:totalder}
\bea
\label{eq:dopeq}
\frac{\d}{\d v(x)}\l\{h^{[0]}_{AL}+\e^2 h^{[2]}_{AL}, f+\e^2 D^{[2]}  f  \r\} &=& o(\e^2)\\
\frac{\d}{\d w(x)}\l\{h^{[0]}_{AL}+\e^2 h^{[2]}_{AL}, f+\e^2 D^{[2]}  f \r\}&=& o(\e^2)
\eea
These two variational equations give rise to an overdetermined linear system of coupled $ODEs$ for the nine $d_{I,l,m}(w)$. Notice that the left hand side is a differential polynomial which is linear in $\de_v^n\de^m_w f$. After enforcing \eqref{eq:linweq}, since $f$ has to be otherwise arbitrary, we can solve the system by imposing vanishing of the coefficient of each monomial $(\de_v^n\de^m_w f)$ $\prod_{j=1}^{k}(v^{(j)}(x))^{I_{2j-1}} (w^{(j)}(x))^{I_{2j}}$ for every $n,m,I$. 
It turns out that the first variational condition \eqref{eq:dopeq} is sufficient to solve for all coefficients. The strategy is to solve the equations starting from the highest non-vanishing value of $n$ (equal to 4 in this case), where the equations are linear algebraic in the coefficients $d$, and then express for lower $n$ all non-differentiated unknowns in terms of the others. With this criterion, the system \eqref{eq:dopeq} boils down to a  second order $ODE$ for a single $d_{n,m,I}$, which following this path of solution turns out to be $d_{2,0,(0,2)}(w)$, plus extra conditions which fully constrain the two constants of integration. The final answer is the one reported in \eqref{eq:dop1loop}. The same method generalizes straightforwardly, albeit resulting considerably heavier from a computational point of view, to the higher orders in $\e$: at $\cO(\epsilon^4)$ this method provides the expressions for $b^{[4]}_{vv}$ and $b^{[4]}_{vvv}$ we reported in \eqref{eq:MISTIMO0}-\eqref{eq:MISTIMO}.


\section{Expansion formulae for hypergeometric functions}

We give here some useful expansion formulae \cite{Kalmykov:2006hu} for the expansion of Gauss' hyergeometric function ${}_2F_1(a,b,c;x)$ around integer values of $a$, $b$, and $c$. By hypergeometric recursions, this can be reduced to the following cases:

\begin{eqnarray}
\label{hypexp1}
&& 
\left.{}_2F_1\left( \begin{array}{c} 1+a_1 \e, 1+a_2\e \\
                 2 + c \e \end{array} \right| z\right) 
= 
\frac{1+c\e}{z}
\Biggl(
- \ln (1-z) 
- \e \Biggl\{
\frac{c-a_1-a_2}{2} \ln^2 (1-z) \nn \\ && \hspace{5mm} + c \Li_{2}(z)
      \Biggr\}
+ \e^2 \Biggl\{
  \left[ (a_1+a_2)c - c^2 - 2 a_1 a_2  \right] \Snp_{1,2}(z) 
+ \left[ (a_1+a_2)c - c^2 - a_1 a_2    \right]  \nn \\
&& \hspace{5mm} \ln(1-z) \Li_{2}(z) 
+ c^2 \Li_{3}(z)
- \frac{1}{6}(c-a_1-a_2)^2 \ln^3 (1-z) 
      \Biggr\}
\nonumber \\ &&  \hspace{5mm}
- \e^3 \Biggl\{
  c \left[ (a_1+a_2)c - c^2 - 2 a_1 a_2 \right] \Snp_{2,2}(z) 
+ c \left[ (a_1+a_2)c - c^2 -   a_1 a_2 \right] \nn \\
&& \hspace{5mm} \ln(1-z) \Li_{3}(z) 
+ (c-a_1) (c-a_2) (c-a_1-a_2)  \bigg[  \ln(1-z) \Snp_{1,2}(z) \nn \\
&& \hspace{5mm}
+ \frac{1}{2}  \ln^2 (1-z) \Li_{2}(z) \bigg]
+ \frac{1}{24} (c-a_1-a_2)^3 \ln^4 (1-z) 
\nn \\
&& \hspace{5mm} + c (c - a_1 - a_2)^2 \Snp_{1,3}(z) 
+ c^3 \Li_{4}(z) 
\Biggr\}
+ {\cO} (\e^4)
\Biggr) \; ,
\label{E-expansion:1}
\end{eqnarray}
\begin{eqnarray}
\label{hypexp2}
&& 
\left. _2F_1\left( \begin{array}{c} a_1 \e, a_2\e \\
                 1 + c \e \end{array} \right| z\right) 
= 
1 + a_1 a_2 \e^2 
\Bigg\{
\Li_{2}(z)
- \e \Biggl[
(c-a_1-a_2) \Snp_{1,2}(z) + c \Li_{3}(z)
      \Biggr]
\nonumber \\ && 
+ \e^2 \Biggl[
 c^2 \Li_{4}(z)
+ (c-a_1-a_2)^2 \Snp_{1,3}(z)
+ \frac{1}{2} \left( c(c-a_1-a_2)+a_1a_2 \right)  \Li_{2}(z)^2
\nonumber \\ && \hspace{10mm}
- \left( c(c-a_1-a_2)+2a_1a_2 \right) \Snp_{2,2}(z)
      \Biggr]
+ {\cO} (\e^3)
\Bigg\} \; .
\label{E-expansion:2}
\end{eqnarray}
In \eqref{E-expansion:1} and \eqref{E-expansion:2}, $\Snp_{n,p}(z)$ is the Nielsen generalized polylogarithm
\beq
\Snp_{n,p}(z) := \frac{(-1)^{n+p-1} }{(n-1)! p!} \int_0^1 \frac{\log ^{n-1}(t) \log ^p(1-t z)}{t} \, dt
\eeq
\end{appendix}
\end{document}